%% file: c-ingestion.tex
\RequirePackage{rotating}
\documentclass[a4paper,fleqn,usenatbib]{mnras}

\usepackage{newtxtext,newtxmath}
\usepackage{rotfloat}
\restylefloat{table}
\usepackage[normalem]{ulem}

\usepackage[T1]{fontenc}
\usepackage{ae,aecompl}

\usepackage{rotating}
\usepackage{graphicx}	
\usepackage{amsmath}	
\usepackage{amssymb}	
\usepackage{booktabs}
\usepackage{natbib}
\usepackage[dvipsnames]{xcolor}
\usepackage{textcomp}
\usepackage{xfrac}
\usepackage{afterpage}

\input{macros/code}




\input{macros/concepts}
\input{macros/vectors}
\input{macros/formatting}

\input{macros/symbols}
\newcommand{\ppmstar}{\code{PPMstar}}
\newcommand{\cldfluid}{$\mathcal{F}_\mrm{C}$}
\newcommand{\airfluid}{$\mathcal{F}_\mrm{O}$}

\newcommand{\MSun}{\mathrm{M}_\odot}
\newcommand{\netone}{\textsc{Net\,1}}
\newcommand{\nettwo}{\textsc{Net\,2}}
\newcommand{\mrm}{\mathrm}
\newcommand{\reactOO}{$^{16}$O+$^{16}$O}
\newcommand{\reactCC}{$^{12}$C+$^{12}$C}
\newcommand{\reactCO}{$^{12}$C+$^{16}$O}

\title[C ingestion into a convective O shell]{3D hydrodynamic simulations of C
ingestion into a convective O shell}

\author[R. Andrassy et al.]{
R. Andrassy,$^{1,2,3}$\thanks{E-mail: andrassy@uvic.ca}
F. Herwig,$^{1,2}$
P. Woodward,$^{4,2}$
C. Ritter$^{1,2}$
\\
$^1$ Department of Physics \& Astronomy, University of Victoria, PO Box 1700
STN~CSC, Victoria, BC, V8W~2Y2, Canada\\
$^2$ Joint Institute for Nuclear Astrophysics, Center for the Evolution of the
Elements, Michigan State University,\\ 640 South Shaw Lane, East Lansing, MI
48824, USA\\
$^3$ Heidelberg Institute for Theoretical Studies, Schloss-Wolfsbrunnenweg 35,
69118 Heidelberg, Germany\\
$^4$ LCSE and Department of Astronomy, University of Minnesota, Minneapolis, MN
55455, USA
}

\date{Accepted XXX. Received YYY; in original form ZZZ}

\pubyear{2018}

\begin{document}
\label{firstpage}
\pagerange{\pageref{firstpage}--\pageref{lastpage}}
\maketitle

\begin{abstract}
Interactions between convective shells in evolved massive stars have been linked
to supernova impostors, to the production of the odd-Z elements Cl, K, and Sc,
and they might also help generate the large-scale asphericities that are known
to facilitate shock revival in supernova explosion models. We investigate the
process of ingestion of C-shell material into a convective O-burning shell,
including the hydrodynamic feedback from the nuclear burning of the ingested
material. Our 3D hydrodynamic simulations span almost 3 dex in the total
luminosity $L_\mrm{tot}$. All but one of the simulations reach a
quasi-stationary state with the entrainment rate and convective velocity
proportional to $L_\mrm{tot}$ and $L_\mrm{tot}^{1/3}$, respectively. Carbon
burning provides $14$\,--\,$33\%$ of the total luminosity, depending on the set
of reactions considered. Equivalent simulations done on $768^3$ and $1152^3$
grids are in excellent quantitative agreement. The flow is dominated by a few
large-scale convective cells. An instability leading to large-scale oscillations
with Mach numbers in excess of $0.2$ develops in an experimental run with the
energy yield from C burning increased by a factor of 10. This run represents
most closely the conditions expected in a violent O-C shell merger, which is a
potential production site for odd-Z elements such as K and Sc and which may seed
asymmetries in the supernova progenitor. 1D simulations may underestimate the
energy generation from the burning of ingested material by as much as a factor
two owing to their missing the effect of clumpiness of entrained material on the
nuclear reaction rate.
\end{abstract}



\section{Introduction}

Stellar evolution models show that evolved massive stars contain complex
structures of nuclear burning shells deep in their cores, where most of the
nuclear and binding energy is concentrated
\citep[e.g.][]{Woosley:2002ck,Davis:2018jz}. Energetic nuclear burning of
different nuclear fuels (the most important being C, Ne, O, Si) often drives
convective shells. The convective flows may entrain fresh fuel from layers above
the shell, which burns and invigorates the convection. Under certain conditions,
this positive feedback may result in a fuel ingestion runaway so rapid and
violent that standard assumptions of 1D stellar evolution theory break down
\citep{herwig14}.

\citet{Smith:2014go} suggest that hydrodynamic instabilities in evolved massive
stars could be responsible for the eruptive mass loss events preceding type IIn
supernovae \citep[SN impostors, see also][]{smith11,ofek14, arcavi17}, which
comprise $6$--$12$\% of all core-collapse supernovae \citep{smith11b}. Such an
instability, if it occurs deep in the core and shortly before the ultimate core
collapse, could also provide the large-scale and large-amplitude seed
perturbations that facilitate shock revival in multidimensional supernova
explosion models \citep{couch_ott13, couch_ott15, muller_janka15, muller17}. The
instability would likely involve chemical mixing between nuclear burning shells
in the core with some impact on element production in the star.

\citet{ritter18}, inspired by a merger of convective C- and O-burning shells in
their stellar evolution model of a $15$\,M$_\odot$ star, investigate the unusual
nucleosynthesis that occurs when a large amount of C-shell material is rapidly
brought into the hot O-shell environment. They conclude, in agreement with
earlier reports by \citet{rauscher02} and \citet{tur07}, that the odd-Z elements
P, Cl, K, and Sc, which are underproduced in current galactic chemical evolution
models, can be synthesised this way. Similarly, \citet{clarkson18} find a highly
energetic H-He convective shell interaction event in their model of a
$45$\,M$_\odot$ Pop~III star when the gap between convective H and He shells
closes. Their single-zone calculations of the resulting i-process
nucleosynthesis can reproduce certain intermediate-mass-element features in the
abundance distribution of the most Fe-poor CEMP-no (C-enhanced metal poor) stars
observed to date. However, all of these implications for stellar nucleosynthesis
should be considered qualitative, because they depend on rather uncertain 1D
models of rapidly evolving convective-reactive flows, which are inherently
three-dimensional.

2D simulations of the violent shell burning that occurs in supernova progenitors
shortly before core collapse had become feasible in the 1990s as demonstrated by
\citet{arnett94}. \citet{bazan_arnett94, bazan_arnett98}, and
\citet{arnett_meakin11} further improved the methodology and ultimately included
in their (still 2D) simulations everything from the upper parts of the Si core
to the H envelope. They reported on convection dominated by large-scale
structures, significant deviations from spherical symmetry as well as hot-spot
burning of convectively entrained $^{12}$C just minutes before core collapse.
More recently, 3D simulations have been constructed \citep{meakin_arnett07,
couch_ott15, jones17, muller17, mocak18}, showing that 2D simulations
significantly overestimate the amplitude of convective motions
\citep{meakin_arnett06}.

\citet[][J17 hereafter]{jones17} presented a set of idealised 3D hydrodynamic
simulations constructed to closely resemble a convective O-burning shell in
their 1D model of a $25$\,M$_\odot$ star. They derive a mass entrainment rate at
the upper convective boundary of $5.4 \times 10^{-7}$\,M$_\odot$\,s$^{-1}$ for
the luminosity of the 1D model. It would only take ${\sim}6\,$days --- less than
the lifetime of the O shell --- for the O-shell convection to reach the bottom
of a neighbouring convective C-burning shell at this rate. Some material from
the C shell could thus be entrained into the hot O-shell environment or the two
shells might even merge, as suggested by the nuclear astrophysics results of
\citet{ritter18}. As a first essential step towards a full merger simulation, we
construct a set of 3D hydrodynamic simulations to investigate the dynamics of C
entrainment from a stable layer into a convective O-burning shell. We intend to
answer questions like: How strong is the feedback from C burning on the flow in
the shell? How does it depend on the O-luminosity of the shell? Is the
entrainment process stable? Does it ever lead to a fuel ingestion runaway such
as that described by \citet{herwig14}? How is the luminosity--entrainment rate
relation measured by J17 affected by C burning?

\section{Methods}

\subsection{1D stellar-evolution model}

The 3D simulations described in this work are based on the 1D evolution model of
a $25\,\MSun$ star computed by J17 using the \mesa{} code \citep{paxton11,
paxton13, paxton15}. The stellar model's initial metallicity is \mbox{$Z =
0.02$}. Rotation is not considered and the Schwarzschild criterion is used to
delineate convective regions. Mixing of chemical species at convective
boundaries is modelled using a diffusion coefficient decaying exponentially with
an e-folding length $\frac{1}{2} f_\mrm{CBM} H_p$, where $f_\mrm{CBM} = 0.022$
is used for all convective boundaries before core C ignition except the bottom
boundaries of burning shells where $f_\mrm{CBM} = 0.005$ is set and $f_\mrm{CBM}
= 0.002$ is used for all convective boundaries after core C ignition.

\begin{figure}
\includegraphics[width=\linewidth]{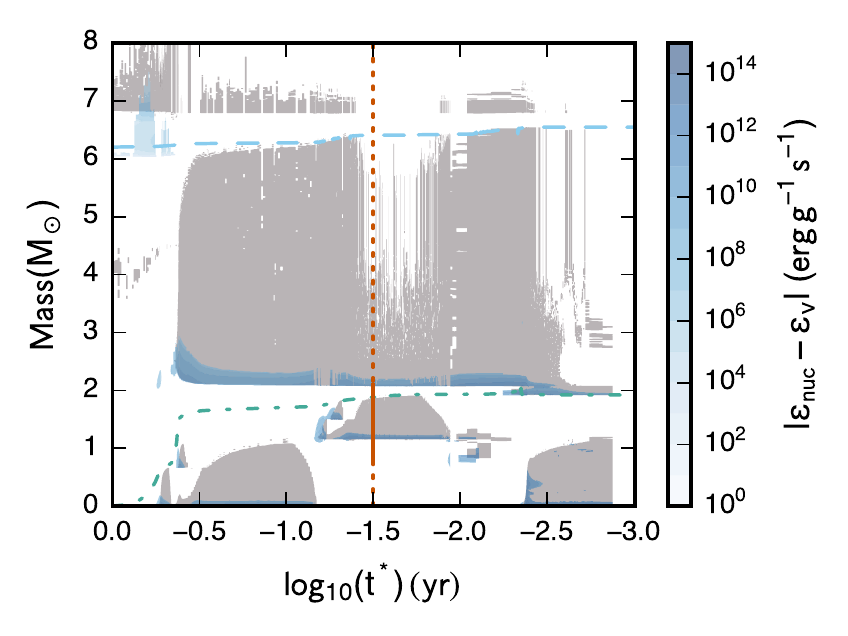}
\caption{Kippenhahn diagram showing the core structure of the $25\,\MSun$
\mesa{} model as a function of time until core collapse. Shades of blue show the
energy generation rate and grey regions are convective. The light-blue dashed
and turquoise dot-dashed lines are the boundaries of the He- and C-free cores,
respectively. The red vertical line indicates the model that was used to
construct the initial condition of the 3D simulations. The solid portion of that
line marks the radial extent of the 3D model.}
\label{fig:kippenhahn}
\end{figure}

We focus on the first shell O-burning phase that spans approximately from 21 to
4 days before core collapse (Fig.~\ref{fig:kippenhahn}). With the start of shell
O burning, a convection zone appears and grows in size until its mass reaches
$0.66\,\MSun$. The outward propagation of the upper convective boundary slows
down at this point and further growth is limited to an additional $0.08\,\MSun$
over $7$\,days, after which convection recedes. At the point of the shell's
maximum extent there is $0.18\,\MSun$ of stable material separating the O shell
from the C shell and there is no mixing between the two in the stellar evolution
model.

\subsection{3D \ppmstar{} simulations}

\subsubsection{\ppmstar{} code}
\label{sect:ppmstar}

We use the \ppmstar{} code of \citet{woodward15}. It is an explicit
Cartesian-grid-based code for 3D hydrodynamics built around the
Piecewise-Parabolic Method \citep[PPM;][]{woodward_colella81,
woodward_colella84, colella_woodward84, woodward86, woodward07}. The code
advects the fractional volume FV of the lighter fluid in a two-fluid scheme
using the Piecewise-Parabolic Boltzmann method \citep[PPB;][]{woodward86,
woodward15}. In 1D, PPB is formally fifth-order accurate, and in our 3D flows,
despite our use of directional splitting, it delivers greatly improved accuracy
over PPM advection. The benefits of PPB advection can be seen in the 2D code
comparison in \citet{joggerst14} and in the 3D code comparison on a
Rayleigh-Taylor problem in \citet{ramaphrabhu12}. PPB achieves its greater
accuracy in part by updating 10 lower-order moments of the distribution of FV
within each cell. With this additional information to work with, it should not
be surprising that we usually find its results to be as good or better than PPM
advection run on a grid with twice the number of grid cells in each dimension.
Also, this additional, subcell information updated by PPB allows us to describe
sharp transitions on the grid from a state with no tracked fluid, FV = 0, to
only tracked fluid, \mbox{FV = 1}, without resorting to the discontinuity
detection and steepening methods of PPM advection. The result is that thin
transitions can be advected with minimal diffusion and also without the changes
in the algorithmic description that can give rise to small glitches with PPM
advection that can be subsequently amplified at physically unstable interfaces.
This feature can be observed in the 3D test problem motivated by inertial
confinement fusion that is discussed in \citet{woodward12}. We also note that
our stellar convection flows involve Mach numbers of the order of $10^{-2}$. It
is therefore especially important that the PPM scheme involves interpolated
estimates of the values of variables at cell edges that are derived from cubic
polynomials. This feature makes PPM well behaved when the scheme is applied at
low Mach numbers, and hence low flow Courant numbers, when these edge values
play an especially important role.

The time step is fixed and corresponds to $\mrm{CFL}=0.75$ at $t=0$.  We impose
constraints where FV approaches the special values of zero and unity.  We do not
constrain nuclear burning rates with the exception of not burning more mass than
is available in the computation cell and not burning trivial amounts of
material. The reader is referred to \citet{woodward07} and \citet{woodward15}
for further details of the overall scheme and of the approximate Riemann solver
implemented in \ppmstar{}. The code was designed with strong emphasis on
parallel efficiency and it has performed past simulations of shell convection on
up to 440,000 CPU cores on the NCSA Blue Waters computer
\citep{woodward15,woodward18,herwig14}.

The version of \ppmstar{} used in this work has four data output channels, all
of which we use in this work:
\begin{itemize}
    \item Full-resolution 3D data cubes. Variable-dependent non-linear
    transforms are applied to compress the data to a single byte per voxel,
    which is sufficient for flow visualisation purposes. The fractional volume
    variable is saved at twice the original grid resolution to make use of the
    PPB subcell information.
    \item 3D data cubes downsampled by averaging blocks of $4^3$ grid cells,
    which decreases spatial resolution by a factor of $4$ and the amount of data
    by a factor of $4^3=64$. Variable-dependent non-linear transforms are
    applied to further compress the data to two bytes per voxel, which is
    sufficient for both flow visualisation and data analysis.
    \item Spherical averages measured in 80 radial, space-filling tetrahedra,
    which we call \emph{buckets} (see Fig.~17 of J17 for their exact
    distribution). The buckets are ideal for rapid analysis of time series
    whenever angular resolution is not critical.
    \item Spherical averages over the full solid angle, which we use to study
    global properties of the flow, mass transport, or burning rates.
\end{itemize}
All of this output is computed on-the-fly and written to disk at regular time
intervals without ever interrupting the hydrodynamic simulation.

\subsubsection{3D simulation setup}
\label{sect:3dsimsetup}

The initial stratification of our simulations is the same as the one used by J17
and is composed of three polytropes that approximate the stratification of the
\mesa\ model. The bottom and top polytropes are stable against convection and
the middle one is adiabatic. The polytropic stratification is close to that of
the \mesa\ model throughout the volume included in the 3D simulation, see Fig.~4
of J17, and little could be gained by modelling the stratification in more
detail given the differences in the equation of state described below. The
polytropic model is also resolved on the 3D computational grid by construction,
which is not true about the \mesa\ model. The stratification changes
significantly on the simulation time scale when we drive the convection strongly
(see Sect.~\ref{sect:luminosity_evolution}), but the change is negligible when
the driving luminosity is close to that of the \mesa\ model. The C shell and
outer layers of the star as well as the inner core are not included.  We impose
boundary conditions at radii of $3.5$\,Mm and $9.2$\,Mm such that the flow
remains confined between the two spherical walls and material fluxes through
them vanish. The boundaries are rough on small scales because of the Cartesian
geometry of the computational grid. This effect is not expected to influence our
results, because we restrict the time interval of our analysis such that the
convective boundary does not get too close to the outer boundary
condition.\footnote{Run I11 is an exception, see Sect.~\ref{sect:stability}.}

We use the equation of state of an ideal monatomic gas in our simulations. This
is a surprisingly good approximation owing to the high temperature typical of
O-shell convection. Despite high density, degeneracy pressure is limited to
$3$--$6$\% of the total pressure in the shell. Its contribution increases to
$\sim 20\%$ in the stable layer below the shell, which we only use as a buffer
zone that forces the convective flows to turn around.  Our two-fluid models (see
below) also cannot include the composition step at the bottom of the shell, so
we refrain from any attempts to quantify mixing at the shell's lower boundary in
this work. Radiation pressure does not exceed $25\%$ of the total pressure in
the shell. A new version of the \ppmstar{} code, which includes radiation
pressure, is already being tested and we intend to quantify the influence of
radiation pressure on stellar convection in a future publication. The
initial stratification of the \ppmstar{} runs is compared with that of the
\mesa{} model in Fig.~4 of J17.

The upper stable layer is initially filled with fluid \cldfluid{} and the rest
is fluid \airfluid. There is a $0.25$\,Mm thick smooth transition in the
fractional volume $\mrm{FV}$ of fluid \cldfluid{} centred at the radius of
$8.08$\,Mm such that $\mrm{FV}=0$ below the transition layer and
\mbox{$\mrm{FV}=1$ above it} (see Fig.~20 of J17). Each of the two fluids
corresponds to a mixture of chemical elements. The individual abundances of
these elements are not evolved in time and only those that enter our simplified
reaction network (see Sect.~\ref{sect:reaction_network}) need to be known. The
concentration of $^{16}$O in fluid \airfluid{}, $X_{16} = 0.382$, is taken from
the \mesa\ model directly. The stable layer above the O shell (fluid
\cldfluid{}) is for the purposes of our experiment assumed to be rich in
$^{12}$C with the mass fraction $X_{12} = 0.13$. This concentration is 5 times
larger than that in the C shell of the \mesa\ model. We do this to speed up the
simulations' transition to a quasi-stationary state. Despite this increase in
$X_{12}$, we use the same mean molecular weights as J17 ($\mu_1 = 1.802$ for
\cldfluid{} and $\mu_2 = 1.848$ for \airfluid) to allow direct comparison of
entrainment rates. The radial profile of the squared Brunt-V\"ais\"al\"a
frequency $N^2$ in this transition layer closely resembles that of the \mesa\
model, see Fig.~6 of J17.

\begin{figure}
\includegraphics[width=0.99\linewidth]{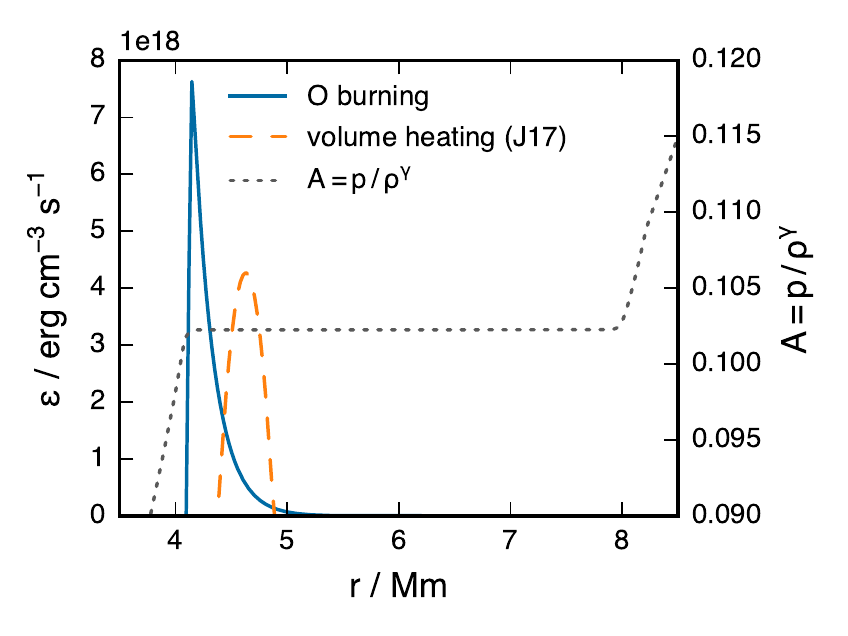}
\caption{Comparison of the distribution of the heating rate per unit volume as
given by the O-burning prescription used in this work with that given by the
volume heating prescription of J17. Both distributions are normalised to give
the same luminosity of $10^{11}\,L_\odot$. The flat part of the entropy profile
$A = p/\rho^\gamma$ is the convection zone.}
\label{fig:heating_distribution}
\end{figure}

J17 used a volume heating term to drive convection. We have implemented an
O-burning prescription in the code (see Sect.~\ref{sect:reaction_network}) and
introduced a parameter ($f_\mrm{OO}$ in Table~\ref{tab:runs}) to scale the heat
output and reach different driving luminosities without having to change the
initial stratification. Figure~\ref{fig:heating_distribution} shows that the
heating rate distribution is more concentrated to the bottom of the shell in our
simulations than in those of J17. Neutrino cooling balances all of the shell's
luminosity in the \mesa{} model. We do not consider this effect and estimate its
influence on the mass entrainment rate at the end of
Sect.~\ref{sect:entrainment_rate}. However, our O-burning prescription with
$f_\mrm{OO} = 1$ (run I2, see Table~\ref{tab:runs}) provides a driving
luminosity $L_\mrm{OO} = 4.27 \times 10^{10}$\,L$_\odot$, which is comparable to
the maximum luminosity $L = 5.2 \times 10^{10}$\,L$_\odot$ reached in the O
shell of the \mesa{} model when both nuclear heating and neutrino cooling are
considered.

\subsubsection{Simulations}
\label{sect:simulations}

\bgroup
\def\arraystretch{1.5}
\begin{sidewaystable*}
\centering
\begin{tabular}{ccccccccccccccc}
\toprule
id &
grid &
$t_\mrm{sim}$ &
nuc. net. &
$f_\mrm{OO}$ &
$f_\mrm{CC}$ &
$f_\mrm{CO}$ &
$L_\mrm{OO}$ &
$L_\mrm{CC}$ &
$L_\mrm{CO}$ &
$L_\mrm{tot}$ &
$\dot{M}_\mrm{e}$ &
$v_\mrm{rms}$ &
$\tau_\mrm{conv}$ &
$\mathrm{Ma}$ \\
 &
 &
[min] &
 &
 &
 &
 &
[L$_\odot$] &
[L$_\odot$] &
[L$_\odot$] &
[L$_\odot$] &
[M$_\odot$\,s$^{-1}$] &
[km\,s$^{-1}$] &
[min] \\
\midrule
D1 & $768^3$ & $55.2$ & --- & --- & --- & --- & --- & --- & --- & $1.18$\,($11$) & $1.15$\,($-6$)
& $38.0$ & $3.46$ & $9.40$\,($-3$) \\
D2 & $1536^3$ & $27.3$ & --- & --- & --- & --- & --- & --- & --- & $1.18$\,($11$) & $1.33$\,($-6$)
& $39.1$ & $3.37$ & $9.70$\,($-3$) \\
D5 & $768^3$ & $37.2$ & --- & --- & --- & --- & --- & --- & --- & $5.91$\,($11$) & $8.07$\,($-6$)
& $66.7$ & $2.05$ & $1.65$\,($-2$) \\
D6 & $768^3$ & $41.3$ & --- & --- & --- & --- & --- & --- & --- & $1.18$\,($12$) & $1.68$\,($-5$)
& $85.8$ & $1.65$ & $2.13$\,($-2$) \\
D8 & $768^3$ & $36.2$ & --- & --- & --- & --- & --- & --- & --- & $2.95$\,($11$) & $3.60$\,($-6$)
& $52.2$ & $2.58$ & $1.29$\,($-2$) \\
D9 & $768^3$ & $43.5$ & --- & --- & --- & --- & --- & --- & --- & $2.95$\,($12$) & $3.83$\,($-5$)
& $118$ & $1.27$ & $2.92$\,($-2$) \\
D10 & $768^3$ & $43.7$ & --- & --- & --- & --- & --- & --- & --- & $5.91$\,($12$) & $7.94$\,($-5$)
& $149$ & $1.03$ & $3.71$\,($-2$) \\
D23 & $768^3$ & $53.4$ & --- & --- & --- & --- & --- & --- & --- & $2.95$\,($10$) & $9.50$\,($-8$)
& $21.0$ & $6.14$ & $5.17$\,($-3$) \\
I2 & $768^3$ & $242$ & \netone & 1 & 1 & --- & $4.27$\,($10$) & $8.30$\,($9$) & --- &
$5.10$\,($10$) & $3.14$\,($-7$)& $27.9$ & $4.74$ & $6.85$\,($-3$) \\
I4 & $768^3$ & $76.3$ & \netone & 2.7 & 1 & --- & $1.16$\,($11$) & $2.84$\,($10$) & --- &
$1.45$\,($11$) & $1.09$\,($-6$)& $40.3$ & $3.34$ & $9.89$\,($-3$) \\
I5 & $768^3$ & $39.5$ & \netone & 13.5 & 1 & --- & $5.79$\,($11$) & $2.54$\,($11$) & --- &
$8.33$\,($11$) & $9.66$\,($-6$)& $117$ & $1.94$ & $1.78$\,($-2$) \\
I11 & $768^3$ & $13.5$ & \netone & 13.5 & 10 & --- & $8.05$\,($11$) &
$1.32$\,($13$) & --- & $1.40$\,($13$) & $2.24$\,($-4$)& $181$ & $0.82$ & $4.46$\,($-2$) \\
I12 & $768^3$ & $46.9$ & \netone & 2.7 & --- & --- & $1.16$\,($11$) & --- & --- & $1.16$\,($11$) &
$7.53$\,($-7$)& $35.3$ & $3.73$ & $8.67$\,($-3$) \\
I13 & $768^3$ & $23.1$ & \netone & 67.5 & 1 & --- & $5.10$\,($12$) & $2.54$\,($12$) & --- &
$7.63$\,($12$) & $1.07$\,($-4$)& $147$ & $1.02$ & $3.63$\,($-2$) \\
I14 & $1152^3$ & $22.4$ & \netone & 67.5 & 1 & --- & $6.26$\,($12$) & $2.87$\,($12$) & --- &
$9.13$\,($12$) & $1.18$\,($-4$)& $163$ & $0.95$ & $4.01$\,($-2$) \\
I15 & $1152^3$ & $58.3$ & \netone & 2.7 & 1 & --- & $1.16$\,($11$) & $3.37$\,($10$) & --- &
$1.50$\,($11$) & $1.37$\,($-6$)& $41.6$ & $3.23$ & $1.03$\,($-2$) \\
I16 & $768^3$ & $73.8$ & \nettwo & 2.7 & 1 & 1 & $1.16$\,($11$) & $2.25$\,($9$) &
$2.10$\,($10$) & $1.39$\,($11$) & $9.73\,(-7)$ & $39.3$ & $3.41$ & $9.66$\,($-3$) \\
I17 & $768^3$ & $27.8$ & \nettwo & 67.5 & 1 & 1 & $4.49$\,($12$) & $3.97$\,($11$) &
$3.59$\,($11$) & $5.25$\,($12$) & $6.30\,(-5)$ &  $139$ & $1.06$ & $3.44$\,($-2$) \\
\bottomrule
\end{tabular}
\vspace{0.5cm}
\caption{Properties of the 3D \ppmstar{} simulations presented in this work. The
first three columns give the run identification code, grid resolution, and the
total length of the simulation. If C burning is included the ``nuc. net.''
column gives which nuclear network was used. Energy output from nuclear
reactions is scaled by factors $f_\mrm{OO}$ (\reactOO{}), $f_\mrm{CC}$
(\reactCC{}), and $f_\mrm{CO}$ (\reactCO{}) and the resulting contributions to
the total luminosity $L_\mrm{tot}$ are $L_\mrm{OO}$, $L_\mrm{CC}$, and
$L_\mrm{CO}$. The last four columns give the entrainment rate $\dot{M}_\mrm{e}$,
rms convective velocity $v_\mrm{rms}$, overturning time scale $\tau_\mrm{conv}$
and a mass-weighted Mach number in the convection zone. The values of
$L_\mrm{OO}$, $L_\mrm{CC}$, $L_\mrm{CO}$, $L_\mrm{tot}$, $\dot{M}_\mrm{e}$,
$v_\mrm{rms}$, $\tau_\mrm{conv}$, and $\mathrm{Ma}$ correspond to averages over
a time window several convective overturns long, which was the same for all
variables but varied from run to run. All runs of the D series except
for D23 were already presented by J17.\vspace{5cm}}
\label{tab:runs}
\end{sidewaystable*}
\egroup

Simulations presented in this work are listed in Table~\ref{tab:runs} along with
some of their global properties. We investigate the luminosity dependence of C
ingestion in the series of runs I2 (O-luminosity enhancement factor $f_\mrm{OO}
= 1$), I4 ($f_\mrm{OO} = 2.7$), I5 ($f_\mrm{OO} = 13.5$), and I13 ($f_\mrm{OO} =
67.5$) done on a $768^3$ grid (see Table~\ref{tab:runs}). C burning was only
turned on at $t = 74$\,min in run I2 by mistake. Runs I14 and I15 are
higher-resolution versions of runs I13 and I4, respectively, done on a $1152^3$
grid to estimate any resolution dependence. We experimentally turned C burning
off in run I12 and enhanced the energy release from C burning by the factor
$f_\mrm{CC} = 10$ in run I11, which are otherwise like runs I4 and I5,
respectively. All the runs mentioned so far use the same C-burning reaction
network \netone{} (see the next section). We investigate how our results depend
on the assumptions about C-burning reactions using runs I16 and I17, which are
like runs I4 and I13, respectively, but they use an alternative reaction network
\nettwo{}. We also use for the analysis runs D1, D2, D5, D6, D8, D9, and D10 of
J17, in which convection is driven by a volume heating term and the burning of
the entrained material is not considered. Finally, we have added run D23 to
extend the D-series of runs towards even lower luminosities.

\subsubsection{Reaction network}
\label{sect:reaction_network}

Convection in the shell is driven by O burning, which we compute using Eq.~18.75
of \citet{kippenhahn12}, neglecting electron screening. We do not model the slow
change in the O mass fraction due to this burning process. The temperature is
slightly overestimated in the shell because of our neglect of radiation pressure
in the equation of state. We correct the temperature profile before computing
energy generation rates using the transformation
\begin{equation}
\vartheta = 1.022\,T_9^{0.64},
\end{equation}
where $T_9$ is the temperature in units of $10^9$\,K and the numerical
coefficients were adjusted until a close fit to the temperature profile of the
\mesa\ model was obtained. In order to model the drop in $T_9$ and $X_{16}$ at
the bottom of the O shell seen in the \mesa\ model, we further modify the
temperature profile using either the transformation
\begin{equation}
\Theta = \frac{1}{2}\{1 - \tanh[200(\vartheta - 2.24)]\}\,\vartheta,
\label{eq:T9_drop_1}
\end{equation}
or the transformation
\begin{equation}
\Theta = \frac{1}{2}\{1.01 - 0.99\tanh[0.07(\rho_3 - 1845)]\}\,\vartheta,
\label{eq:T9_drop_2}
\end{equation}
where $\rho_3$ is the density in units of $10^3$\,g\,cm$^{-3}$.
Equation~\ref{eq:T9_drop_1} is used in low- and medium-luminosity runs I2, I4,
I5, I12, I15, and I16, but the shallowness of the temperature profile in the
lower stable layer makes $\Theta$ too sensitive to the relatively large changes
in the stratification we see in high-luminosity runs.
Equation~\ref{eq:T9_drop_2} is much less sensitive to such
changes.\footnote{Equation~\ref{eq:T9_drop_2} also makes sure that $\Theta$
never drops to zero, which would cause numerical problems in the energy
generation module if a negligible but non-zero amount of the C-rich fluid got
below the convection zone and the energy generation rate calculation was
executed. This sometimes happens when convection becomes too vigorous.} We use
it in runs I11, I13, I14, and I17.

There would be a large number of nuclear reactions involved in the burning of
the ingested material in an actual star. However, the purpose of this study is
not to replicate in detail a shell interaction found in a particular stellar
evolution model. Instead, we aim to study the general behaviour of the ingestion
process when the burning of the ingested material feeds back onto the convective
flow. We use two simple reaction networks based on our earlier 1D calculations
with a large network \citep{ritter18}:
\begin{itemize}
    \item \netone: $^{12}$C($^{12}$C,\,$\alpha$)$^{20}$Ne followed by
    $^{16}$O($\alpha$,\,$\gamma$)$^{20}$Ne. The rate of the former reaction is
    computed from the rate of \reactCC{} given by \citet{caughlan_fowler88} with
    the neutron and proton branches subtracted according to the branching ratios
    used by \citet{dayras77} and \citet{pignatari13}. The $\alpha$ particle is
    assumed to be immediately captured in the latter reaction. The Q values of
    the two reactions, $4.62$\,MeV and $4.73$\,MeV, respectively, are summed.
    \item \nettwo: All channels (n, p, $\gamma$) of \reactCC{} and \reactCO{}.
    The rates are taken from \citet{caughlan_fowler88} and the Q values are
    $3.19$\,MeV and $5.26$\,MeV, respectively. The subsequent reactions induced
    by the released neutrons and protons are not considered for simplicity as
    many temperature-dependent reaction paths are possible. 
\end{itemize}

The rate of the \reactCC{} reaction scales with the square $X_{12}^2$ of the
concentration of $^{12}$C and it is thus important when the concentration is
high (i.e.\ when the entrainment rate is high). On the other hand, the
\reactCO{} reaction becomes important at low concentrations (i.e. low
entrainment rates) owing to the reaction's being linear in $X_{12}$.  Having two
reaction networks allows the reader to judge how different the outcome is when
the entrained material only reacts with itself (\netone{}) compared to a
situation in which reactions with the other fluid are also possible (\nettwo{}).
Additionally, we used the \textsc{PPN} \citep{pignatari:16, ritter18b} code to
confirm that our neglect of electron screening is justified, because it is a
$10$-$20\%$ effect for the range of densities and temperatures relevant to the O
shell problem.

Our implementation of nuclear burning in the two-fluid approximation is as
follows. The fractional volumes $\mrm{FV}$ of fluid \cldfluid{} and $1-\mrm{FV}$
of fluid \airfluid{} are first converted into molar fractions $Y_\mrm{C}$ and
$Y_\mrm{O}$ of the $^{12}$C and $^{16}$O reactants, taking into account that
only the fractions $f_\mrm{C} = 0.187$ and $f_\mrm{O} = 0.540$ of all nuclei in
each of the two fluids represent the reactants.\footnote{We make use of the fact
that the fractional volume of fluid \cldfluid{} entrained into the O shell is
small, so we only need to consider $^{16}$O nuclei present in the dominant fluid
\airfluid{}.} This reflects our mental picture that each of the fluids has some
internal composition, which is not evolved.  A change $\mrm{d}Y_\mrm{C}$ in the
molar fraction $Y_\mrm{C}$ of $^{12}$C is then computed using the equations of
nuclear burning \citep[see e.g.\ Chapter~18 of][]{kippenhahn12} with the rates
described above. Whenever a molar fraction $\mrm{d}Y_\mrm{C}$ of $^{12}$C nuclei
needs to be removed from fluid \cldfluid{}, we are forced to remove the molar
fraction $\mrm{d}Y_\mrm{C}/f_\mrm{C}$ of fluid \cldfluid{}, because the internal
composition of this fluid cannot change in our two-fluid model. The burning
process also cannot change the total density of the two-fluid mixture, so the
amount of fluid \cldfluid{} burnt is replaced by an equivalent amount of fluid
\airfluid{}. While this qualitatively captures the increase in the mean
molecular weight $\mu$ upon nuclear fusion, we would need to introduce more
fluids into the model to make the change in $\mu$ quantitatively correct.  The
heat released from the nuclear reactions is added after every 1D pass to
maximise the temporal accuracy of the numerical scheme.

\section{Results}

\subsection{C entrainment and burning}
\label{sect:c_entrainment}

\begin{figure*} \centering
\includegraphics[width=\linewidth]{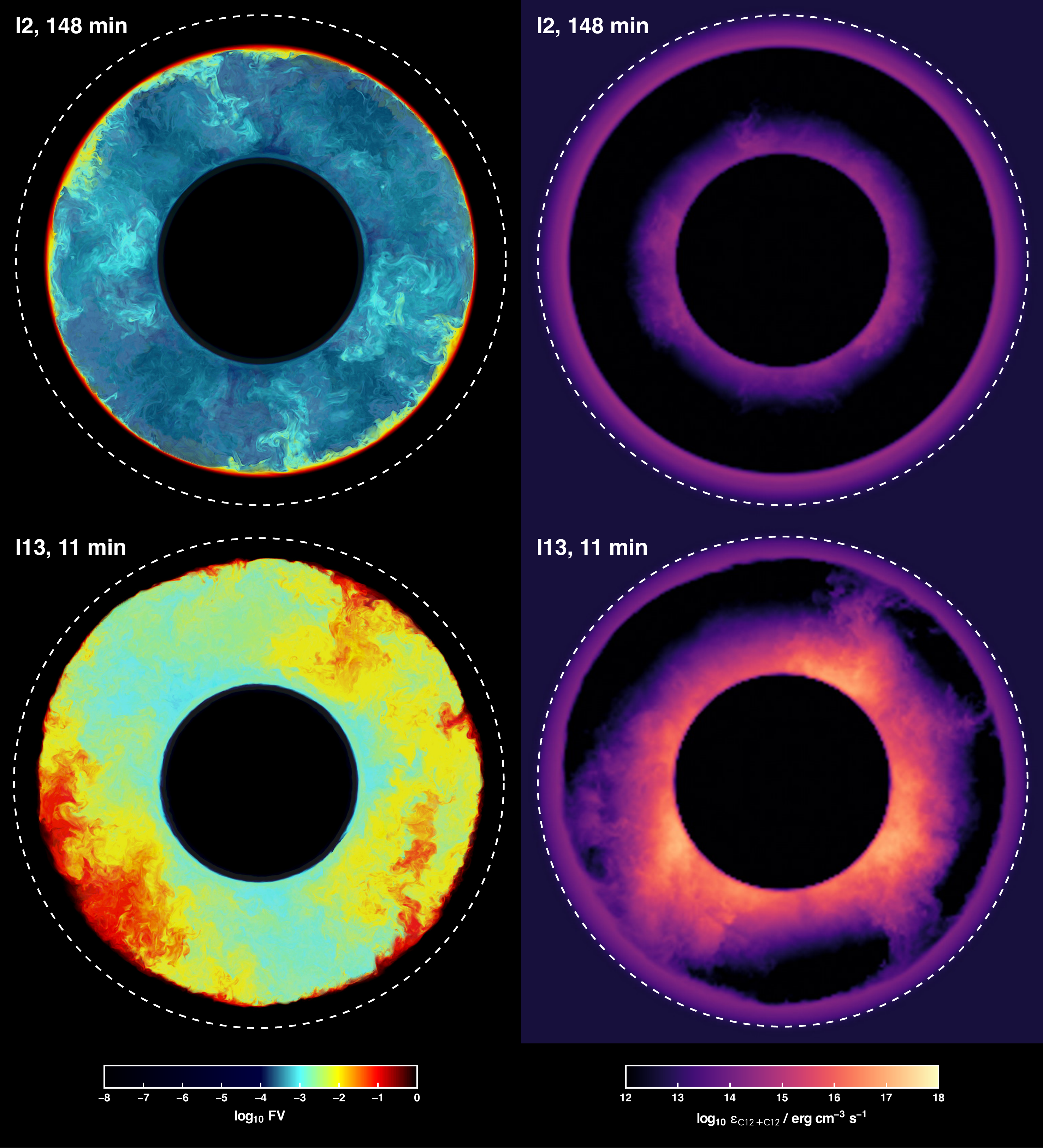}
\caption{Renderings showing a thin slice through the computational box in terms
of the fractional volume FV of fluid \cldfluid{} (left panels) and energy
generation rate from the \reactCC{} reaction (right panels) in runs I2
($f_\mrm{OO} = 1$) and I13 ($f_\mrm{OO} = 67.5$). The two runs differ by a
factor of ${\sim}170$ in the total luminosity at the points in time
corresponding to the renderings. The energy generation rates were computed by
post-processing data downsampled from $768^3$ to $192^3$, in which all 3D
fluctuations except those in FV were neglected, see
Sect.~\ref{sect:fluctuations}. The white dashed circles correspond to the outer
boundary condition.}
\label{fig:renderings_I2_I13}
\end{figure*}

The flow field in our simulations is dominated by a few large-scale convection
cells (Fig.~\ref{fig:renderings_I2_I13}). They cause shear flows as they turn
over at the upper convective boundary, but, as \citet{woodward15} and J17
describe in detail, entrainment does not occur where the shear is the strongest.
It rather occurs at places where two neighbouring flows sliding along the
boundary collide and are forced back into the convection zone by the global flow
topology, dragging slivers of fluid \cldfluid{} along.

According to our assumption, the concentration of $^{12}$C in fluid \cldfluid{}
is five times higher than that in the stellar evolution model (see
Sect.~\ref{sect:3dsimsetup}), so we can see some energy generation due to
\reactCC{} reactions in the upper stable layer in \Fig{renderings_I2_I13}, but
the burning is not strong enough to establish a new convection zone on the time
scales considered. \reactCO{} reactions do not contribute in the stable layer,
because we only compute reactions involving $^{12}$C from fluid \cldfluid{} and
$^{16}$O from fluid \airfluid{} and there is no fluid \airfluid{} in the upper
stable layer. The concentration of \cldfluid{} drops by orders of magnitude at
the upper boundary as the mixing process reduces the buoyancy of the fluid
mixture to make it possible for convection to pull it to the bottom of the
convection zone. Therefore, C-burning reactions are virtually absent in the
upper half of the convection zone and they are only rekindled at the bottom
owing to an increase in temperature.

\begin{figure}
\includegraphics[width=\linewidth]{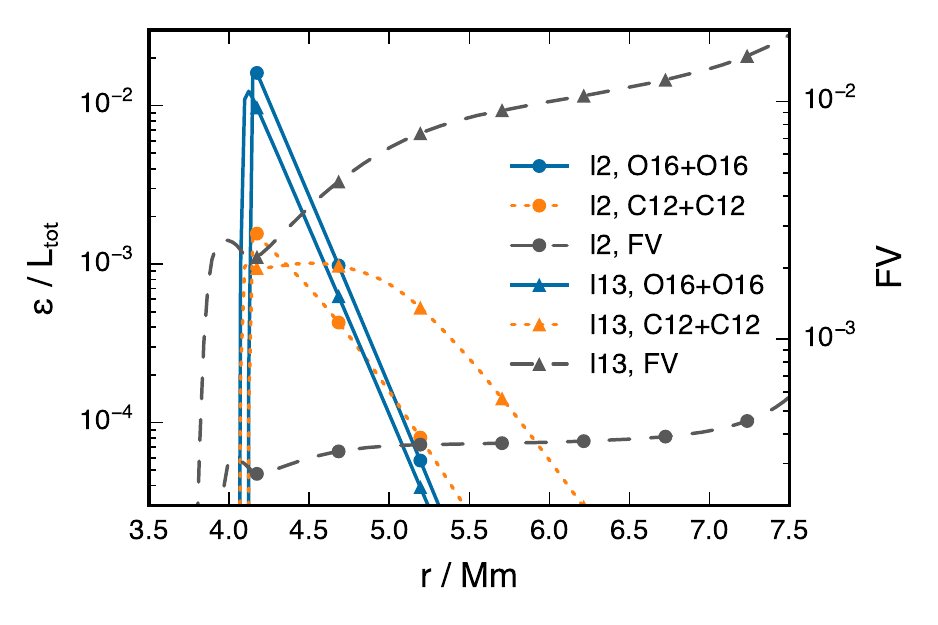}
\caption{Comparison of the energy generation rates per unit volume in the
low-luminosity run I2 ($f_\mrm{OO} = 1$) with those in the high-luminosity run
I13 ($f_\mrm{OO} = 67.5$). The energy generation rate $\varepsilon$ is
normalised by the total luminosity $L_\mrm{tot}$ and the radial profiles of the
fractional volume FV of the entrained fluid are shown for comparison. The radial
profiles correspond to $148$\,min in I2 and to $11$\,min in I13 with time
averaging over two convective overturning time scales applied around these
points in time in both cases.}
\label{fig:energy_sources_I2_I13}
\end{figure}

Because the \reactCC{} reaction rate depends on the square of the concentration
$X_{12}$ of $^{12}$C, the burning time scale in runs with a significant
contribution from the \reactCC{} reaction becomes shorter when the driving
luminosity and hence also the mass entrainment rate is increased. The burning
time scale is a few times longer than the convective overturning time scale in
the low-luminosity run I2 ($f_\mrm{OO} = 1$), which results in a rather flat
fractional volume profile of the entrained C-rich fluid, see
Fig.~\ref{fig:energy_sources_I2_I13}. On the other hand, there is a significant
fractional volume gradient in the convection zone of the high-luminosity run I13
($f_\mrm{OO} = 67.5$), in which the two time scales are comparable.  This
gradient causes the C-burning layer to be more extended in run I13 than in run
I2 (Fig.~\ref{fig:energy_sources_I2_I13}). The burning time scale for the
\reactCO{} reaction is independent of $X_{12}$.

Figure~\ref{fig:renderings_I2_I13} also shows that there is a lot of large scale
structure in the distribution of the entrained material. We show in
Sect.~\ref{sect:fluctuations} that these inhomogeneities are the main cause of
the asymmetric distribution of the burning rate
(Fig.~\ref{fig:renderings_I2_I13}) with temperature fluctuations playing a
secondary role. The asymmetry is the most pronounced at the upper end of the
luminosity range considered.

All runs, with the  exception of run I11, reach a quasi-stationary state in
which there is a close balance between mass entrainment and burning and the
concentration of fluid \cldfluid{} in the convection zone stays approximately
constant. Some selected properties of our simulations are summarised in
Table~\ref{tab:runs}. The following three sections present the detailed
evolution of the simulations in terms of the entrainment rate, luminosity, and
velocity field. We analyse 3D fluctuations of the most relevant quantities in
Sect.~\ref{sect:fluctuations} and, finally, we discuss the unstable run I11 in
Sect.~\ref{sect:stability}.

\subsection{Entrainment rate}
\label{sect:entrainment_rate}

The mass $M_\mrm{e}(t)$ of the C-rich fluid \cldfluid{} entrained into the
convection zone by time $t$ is the sum
\begin{equation}
M_\mrm{e}(t) = M_\mrm{p}(t) + M_\mrm{b}(t)
\end{equation}
of the mass $M_\mrm{p}(t)$ of \cldfluid{} present in the convection zone at time
$t$ and the mass $M_\mrm{b}(t)$ of \cldfluid{} burnt in the convection zone by
time $t$.

To compute $M_\mrm{p}(t)$, we employ the method of J17. We first determine the
radius $r_\mrm{ub}(t)$ of the upper boundary of the convection zone as defined
by the position of the steepest decline in the spherically-averaged rms
tangential velocity $v_\perp$, and the radius $r_\mrm{top}(t) = r_\mrm{ub}(t) -
H_\mrm{v,ub}(t)$, where $H_\mrm{v,ub}(t) = v_\perp|\partial v_\perp / \partial
r|^{-1}$ is the scale height of $v_\perp$ to be evaluated at $r_\mrm{ub}(t)$.
$M_\mrm{p}(t)$ is then the volume integral of the density of fluid \cldfluid{}
inside the radius $r_\mrm{top}(t)$.\footnote{The amount of fluid \cldfluid{}
getting below the convection zone is negligible.}

The burnt mass $M_\mrm{b}(t)$ is given by a volume and time integral of the mass
burning rate in the convection zone, which is computed from reaction rates in a
way analogous to the luminosity computation described in
Sect.~\ref{sect:luminosity_evolution}. Since the burning is concentrated to the
lower part of the convection zone, we simply integrate up to the radius of
$7.5$\,Mm, which is slightly below the initial location of the upper boundary of
the convection zone.

\begin{figure*}
\begin{minipage}{.5\textwidth}
  \centering
  \includegraphics[width=\textwidth]{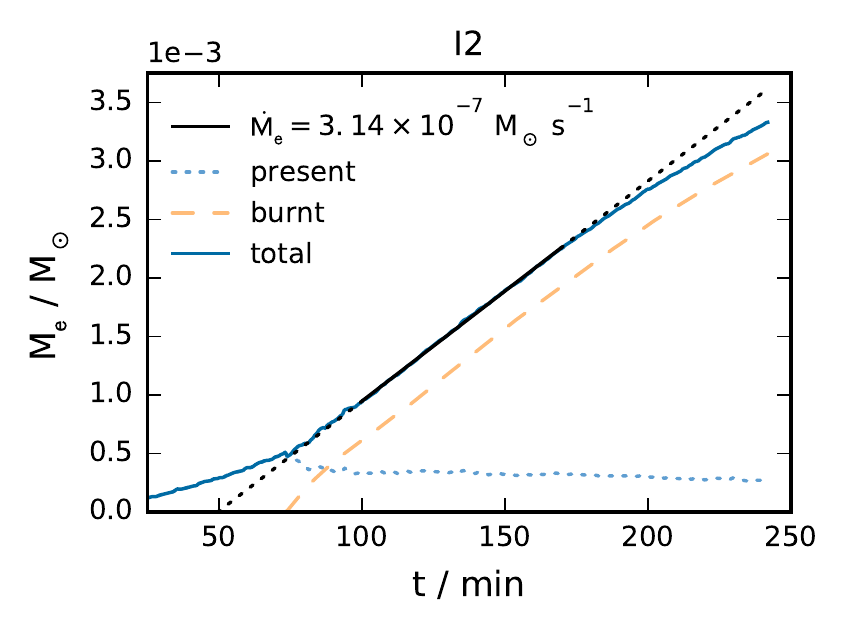}
\end{minipage}%
\begin{minipage}{.5\textwidth}
  \centering
  \includegraphics[width=\textwidth]{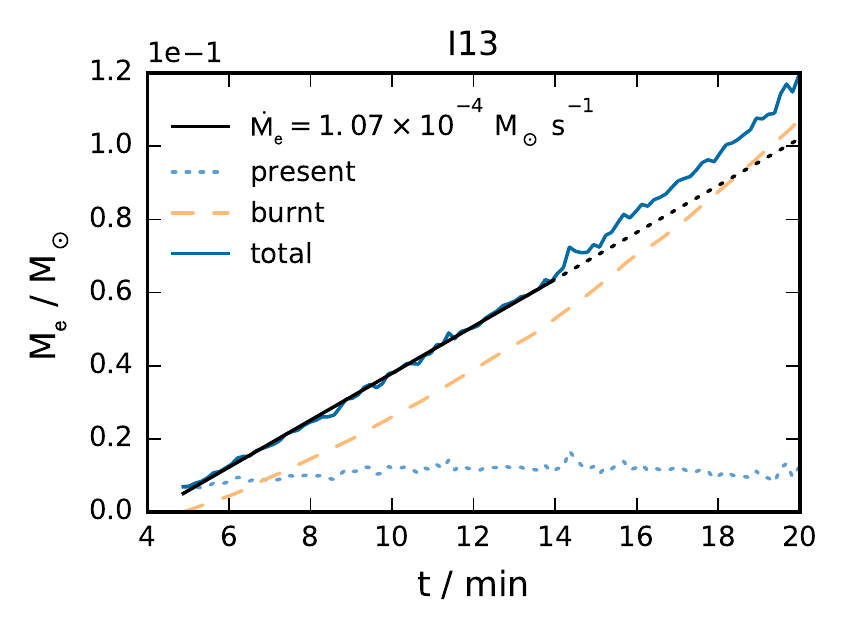}
\end{minipage}
\caption{Total mass of fluid \cldfluid{} that has been entrained into the
convection zone is the sum of the mass of \cldfluid{} currently present in the
convection zone and the mass of \cldfluid{} that has been burnt in the
convection zone. The rate at which the total mass of \cldfluid{} in the
convection zone increases with time is the entrainment rate $\dot{M}_\mrm{e}$.
In each panel, a linear fit to $M_\mrm{e}(t)$ is computed in the time interval
where the black fitting line is solid; the dotted parts are to guide the eye.
The deviations from the fitting line are discussed in
Sect.~\ref{sect:entrainment_rate}. The luminosity driving convection is ${\sim}
150\times$ higher in run I13 (($f_\mrm{OO} = 67.5$), right panel) as compared
with I2 (($f_\mrm{OO} = 1$), left panel), see Table~\ref{tab:runs}.}
\label{fig:entrainment_rate}
\end{figure*}

Figure~\ref{fig:entrainment_rate} shows the entrained mass $M_\mrm{e}(t)$ for
runs I2 ($f_\mrm{OO} = 1$) and I13 ($f_\mrm{OO} = 67.5$), which are the least
and most luminous of the $768^3$ I-series runs with quasi-stationary C burning,
respectively. In run I2, $M_\mrm{p}(t)$ steadily increases until C burning is
turned on at $t = 74$\,min (see Sect.~\ref{sect:simulations}), after which it
takes $3$--$5$ convective overturns to reach a quasi-stationary state with an
entrainment rate of $\dot{M}_\mrm{e} = 3.14 \times
10^{-7}$\,M$_\odot$\,s$^{-1}$. $\dot{M}_\mrm{e}$ decreases again very late in
the run, which is likely caused by the steepening of the entropy gradient at the
upper convective boundary due to the energy release from C burning at the bottom
of the upper stable layer (see Sect.~\ref{sect:c_entrainment}). This effect is
negligible in run I13 (Fig.~\ref{fig:entrainment_rate}), which is $12 \times$
shorter than I2 but, owing to the strong driving of convection, it reaches an
entrainment rate so high that all of the upper stable layer is ultimately
entrained within the simulated time. The quasi-stationary entrainment rate is
$\dot{M}_\mrm{e} = 1.07 \times 10^{-4}$\,M$_\odot$\,s$^{-1}$ in I13. In light of
the linear dependence of the entrainment rate on the total luminosity discussed
in the next paragraph, it is surprising that the entrainment rate is essentially
constant while the total luminosity more than doubles between $5$ and $14$\,min
of simulation time, see Fig.~\ref{fig:luminosity_evolution}. This might be due
to the following opposing effect: as the convective boundary moves further into
the stable layer, the entropy jump across the boundary increases, which hinders
mass entrainment.  The entrainment rate only starts to increase at $t \approx
14$\,min when the convective boundary has reached the radius of $8.6$\,Mm
(starting from $8.0$\,Mm at $t = 0$) and half of the upper stable layer (in
terms of radius) has been engulfed by the convection. The scale height
$H_\mrm{v,ub}$ of the velocity profile starts increasing at that point. We see
the same effect in similarly luminous runs I14 ($f_\mrm{OO} = 67.5$, \netone{}),
I17 ($f_\mrm{OO} = 67.5$, \nettwo{}), and also in run D10 that does not include
C burning, although it only occurs when the boundary has reached $8.75$\,Mm in
D10. A likely explanation is that the outer boundary condition at $9.2$\,Mm
starts to influence the flows at the top of the convection zone when the upper
stable layer has become too thin. The last three minutes of run I13 are not
shown in Fig.~\ref{fig:entrainment_rate}, because velocity amplitudes in the
upper stable layer become so large in that time interval that the
velocity-gradient-based method of locating the convective boundary becomes
unusable. We also exclude from the entrainment analysis the last few minutes of
runs I14, I17, and D10 for the same reason.

\begin{figure}
\includegraphics[width=\linewidth]{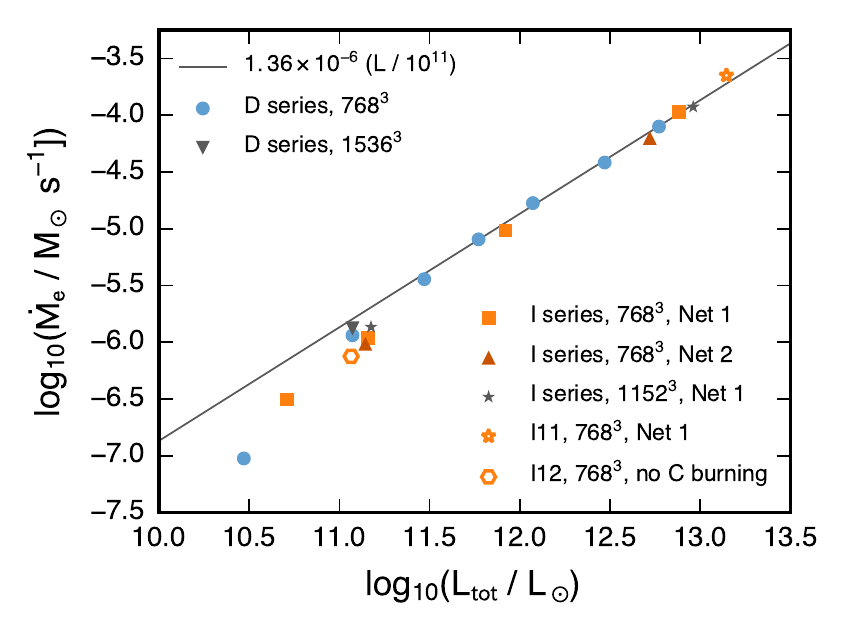}
\caption{Dependence of the entrainment rate on the total luminosity
$L_\mrm{tot}$ integrated over the convection zone. A linear scaling law
essentially identical to the one given in J17 is shown for comparison. The last
$2.2$\,min of run I11 are used when the instability described in
Sect.~\ref{sect:stability} has fully developed.}
\label{fig:entrainment_rate_vs_luminosity}
\end{figure}
The I-series runs with either C-burning network confirm the linear
relation between the entrainment rate and the total luminosity established by
the D-series runs of J17, see Fig.~\ref{fig:entrainment_rate_vs_luminosity}.
Run I11 is close to the scaling relation despite its unstable nature (see
Sect.~\ref{sect:stability} for details). That the entrainment rates
in low-luminosity runs fall below the linear trend is likely caused by the
limited length of our simulations. One would ideally want to run a simulation
of convective boundary mixing at least until all of the initial transition
layer between the two fluids has been entrained and a new boundary has been
formed, consistent with the properties of the convective flows and of the
entrainment process. However, this is very expensive to achieve even on a
$768^3$ grid when the luminosity is low. The initial transition layer at the
upper boundary contains $2.7\times 10^{-2}$\,M$_\odot$ of fluid \cldfluid{},
but only $3.3\times 10^{-3}$\,M$_\odot$ was entrained in the low-luminosity run
I2 ($5.1\times 10^{10}$\,L$_\odot$, see Fig.~\ref{fig:entrainment_rate}), which
involved $3.8\times 10^6$ time steps. The entrainment rate may depend on the
assumed structure of the transition layer in such cases. On the other hand,
most of the initial transition layer was entrained in runs D5 ($5.9\times
10^{11}$\,L$_\odot$, see Fig.~A1 of J17) and I5 ($8.3\times
10^{11}$\,L$_\odot$, see Fig.~\ref{fig:entrainment_rate_appx}) and the layer
was completely engulfed by convection in all runs with $L_\mrm{tot}
\geq 10^{12}$\,L$_\odot$.

Some fraction of the entrainment rates reported here is a consequence of the
gradual increase in the mean entropy of the convection zone, which results from
our neglect of neutrino cooling. The rate of convective boundary migration due
to this effect is proportional to the rate of change of entropy $A =
p/\rho^\gamma$ in the convection zone and inversely proportional to the slope of
$A(m)$ in the relevant part of the stable layer, where $m$ is the enclosed mass
coordinate. During the time interval in which we determined the entrainment rate
$\dot{M}_\mrm{e} = 1.07 \times 10^{-4}$\,M$_\odot$\,s$^{-1}$ in run I13 (see
Fig.~\ref{fig:entrainment_rate}), we measure $(\partial A / \partial
t)_\mrm{conv} = 7.3 \times 10^{-6}$\,s$^{-1}$ in the convection
zone\footnote{This value changes by about a factor of two reflecting the
luminosity increase during the time interval considered. We use the mean value
in this estimate and remind the reader that $\dot{M}_\mrm{e}$ does not change in
the same time interval, see Fig.~\ref{fig:entrainment_rate}. The contribution to
$(\partial A / \partial t)_\mrm{conv}$ from mass entrainment is small and
essentially the whole rate can be explained by energy generation in the
convection zone.}. The slope $(\partial A / \partial m)_\mrm{stab} = 1.4 \times
10^{-1}$\,M$_\odot^{-1}$ fits well the stable stratification above the
convection zone at $t = 0$. As the entropy in the convection zone increases,
stable material is rendered unstable at the rate $\dot{M}_A = (\partial A /
\partial t)_\mrm{conv}\,/\,(\partial A / \partial m)_\mrm{stab} = 5.2 \times
10^{-5}$\,M$_\odot$\,s$^{-1}$, which accounts for $49\%$ of the entrainment rate
in this run. This fraction should not depend on the driving luminosity
$L_\mrm{tot}$, because both $(\partial A / \partial t)_\mrm{conv}$ and
$\dot{M}_\mrm{e}$ scale in proportion to $L_\mrm{tot}$. The estimate just
presented excludes the effect of neutrino cooling on convective velocity.
\citet{Meakin:2007dj} include neutrino cooling in their 3D simulation of a
similar O shell and the radial velocity in their simulation is indeed lower than
what our scaling law predicts (see Fig.~13 of J17). Nevertheless, their mass
entrainment rate is in very good agreement with our scaling law (see Fig.~19 of
J17) despite some differences in the initial stratification.

\subsection{Luminosity evolution}
\label{sect:luminosity_evolution}

We calculate the global burning rates using spherically-averaged profiles of
density, temperature, and fractional volume of fluid \cldfluid{} in a
post-processing step. The spherically-averaged rate of the \reactCC{} reaction
scales with $\langle X_{12}^2 \rangle = \langle X_{12} \rangle^2 +
\sigma_{12}^2$, where $\sigma_{12}^2$ is the standard deviation of the
distribution of $X_{12}$ and the distribution is taken over the full solid
angle at a constant radius. We first compute the rate of \reactCC{} using
$\langle X_{12} \rangle^2$ and then multiply the result by the factor $\xi = 1
+ \sigma_{12}^2 / \langle X_{12} \rangle^2$, where $\sigma_{12}$ is computed
from the standard deviation of the fractional volume of fluid \cldfluid{}. In
this way, we take into account the presence of $^{12}$C clumps in the
convection zone, which increases the resulting luminosity by a factor ranging
from ${\sim}1.2$ in the least luminous runs to ${\sim} 2$ in the most luminous
ones compared to the luminosity calculated using the spherical average $\langle
X_{12} \rangle$ only. The presence of clumps is a 3D
effect and as such it is neglected in 1D stellar evolution calculations. The
non-linear scaling of the \reactOO{} reaction with the mass fraction $X_{16}$
of $^{16}$O is inconsequential as the fractional volume of fluid \airfluid{}
deep in the convection zone is essentially unity at all times.
\begin{figure*}
\begin{minipage}{.5\textwidth}
  \centering
  \includegraphics[width=\textwidth]{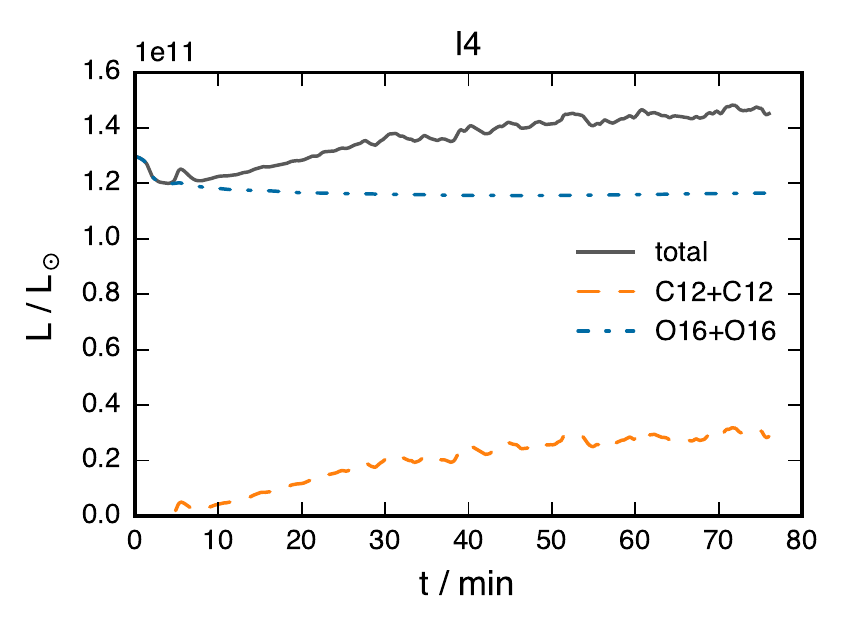}
\end{minipage}%
\begin{minipage}{.5\textwidth}
  \centering
  \includegraphics[width=\textwidth]{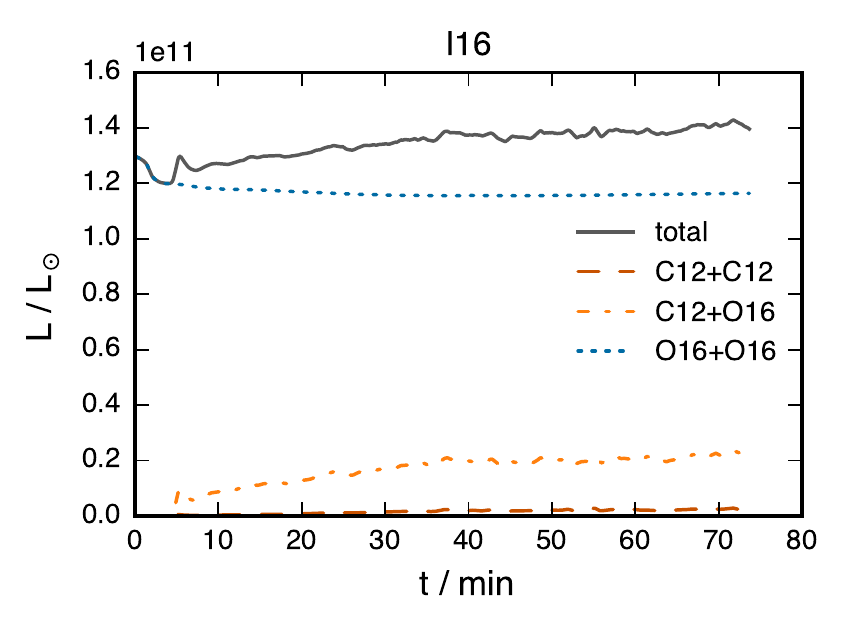}
\end{minipage}\\
\begin{minipage}{.5\textwidth}
  \centering
  \includegraphics[width=\textwidth]{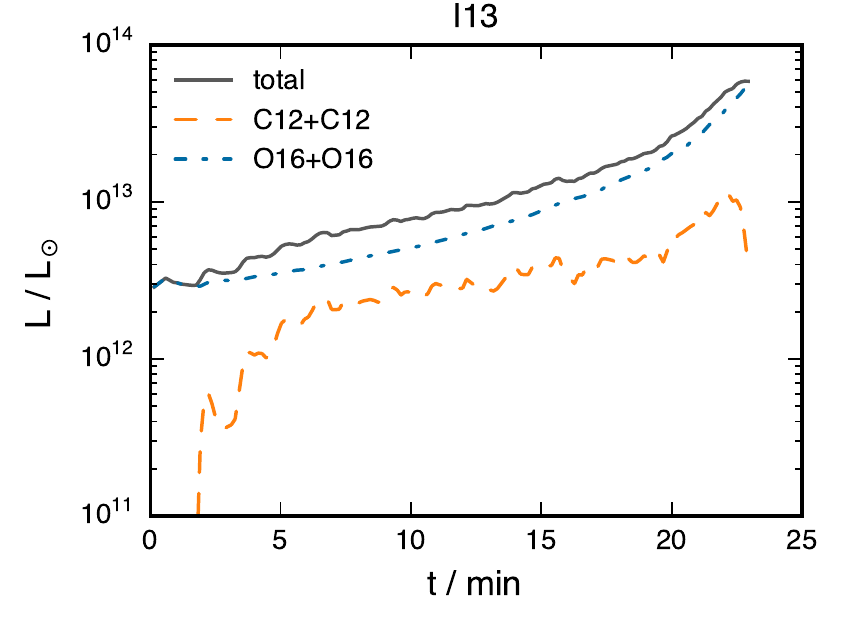}
\end{minipage}%
\begin{minipage}{.5\textwidth}
  \centering
  \includegraphics[width=\textwidth]{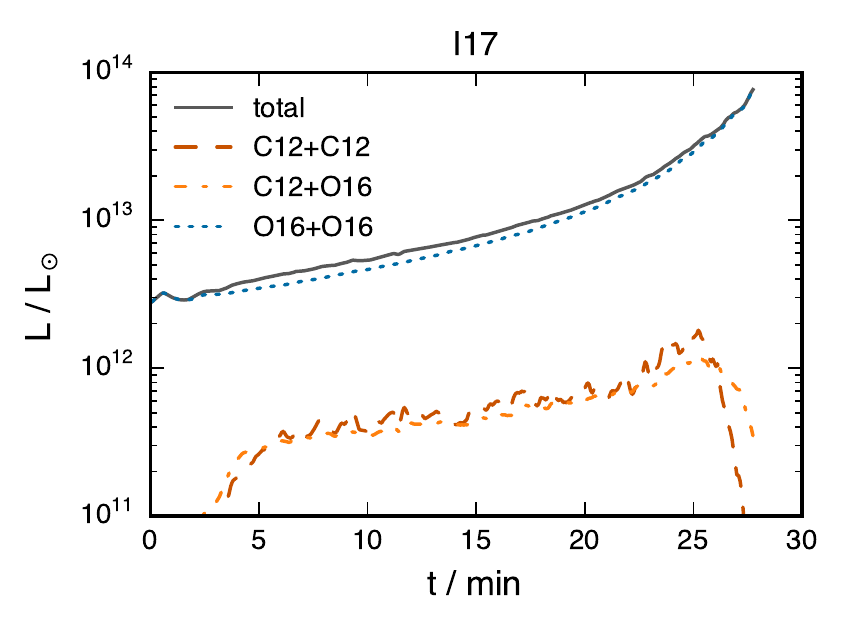}
\end{minipage}
\caption{Time dependence of the contributions to the total luminosity integrated
over the convection zone. The left two panels show runs I4 ($f_\mrm{OO} = 2.7$)
and I13 ($f_\mrm{OO} = 67.5$) with the burning network \textsc{Net1}
(\reactCC{}) and the two panels on the right shown runs I16 ($f_\mrm{OO} = 2.7$)
and I17 ($f_\mrm{OO} = 67.5$), which use \textsc{Net2} (\reactCC{} and
\reactCO{}, see Sect.~\ref{sect:reaction_network} for details).}
\label{fig:luminosity_evolution}
\end{figure*}

The average luminosities are computed for the time intervals during which the
entrainment rates were measured (see Figs.~\ref{fig:entrainment_rate},
\ref{fig:entrainment_rate_appx}, and \ref{fig:entrainment_rate_appx2}), and they
are summarized in Table~\ref{tab:runs}. The time evolution of the luminosity
contributions from O and C burning is shown in
Fig.~\ref{fig:luminosity_evolution} for four characteristic cases. In the
low-luminosity ($f_\mathrm{OO} = 2.7$) run I4, the O-burning luminosity first
decreases during the initial adjustment of the stratification and then it levels
off. The C-burning luminosity (\reactCC{}) increases until an equilibrium is
achieved between the rates of entrainment and burning with C burning providing
${\sim} 20\%$ of the total luminosity. The I16 case has the same O luminosity
but it employs the C-burning network \nettwo{}, which includes the \reactCO{}
reaction as well. This reaction dominates C burning when the concentration of
$^{12}$C in the convection zone is low as is the case in the low-luminosity runs
I4 and I16. The contribution of C burning to the total luminosity is ${\sim}
17\%$ in I16.

Carbon burning is responsible for ${\sim} 33\%$ of the total luminosity in the
more luminous run I13 ($f_\mathrm{OO} = 67.5$, \netone{}), but this
fraction drops to ${\sim} 14\%$ when the reaction \reactCO{} is included in the
otherwise similar run I17 ($f_\mathrm{OO} = 67.5$, \nettwo{}), see
Fig.~\ref{fig:luminosity_evolution}. The reactions \reactCC{} and \reactCO{} are
equally important in I17.

\begin{figure}
\includegraphics[width=\linewidth]{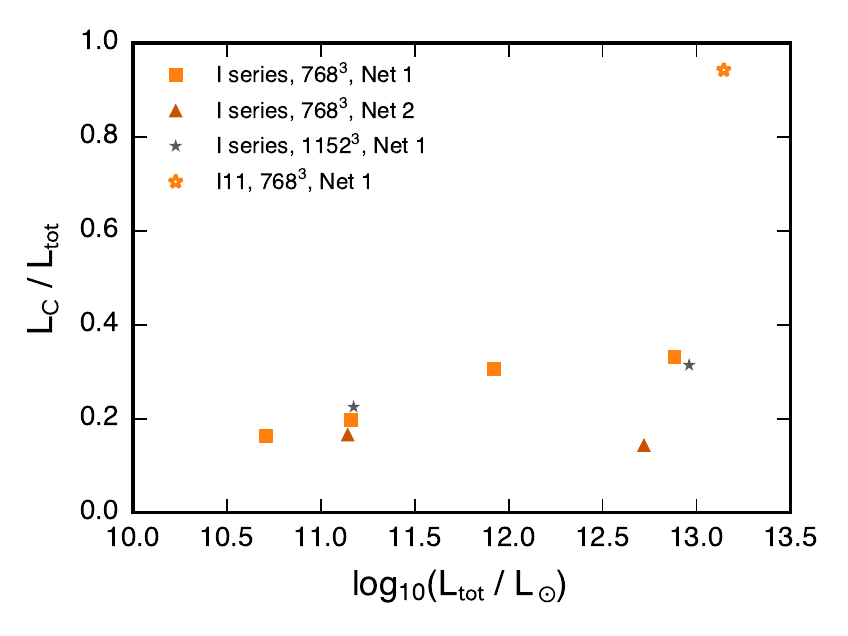}
\caption{Dependence on the total luminosity of the fraction $L_\mrm{C}
/ L_\mrm{tot}$, where $L_\mrm{C}$ is the luminosity from C-burning reactions and
$L_\mrm{tot}$ the total luminosity, both integrated over the convection zone.
The last $2.2$\,min of run I11 are used when the instability described in
Sect.~\ref{sect:stability} has fully developed.}
\label{fig:L_C_fraction_vs_luminosity}
\end{figure}

The luminosity $L_\mrm{C}$ from the C-burning reactions makes up an increasingly
larger fraction of the total luminosity $L_\mrm{tot}$ as $L_\mrm{tot}$ increases
from run to run when burning network \netone{} is used, see
Fig.~\ref{fig:L_C_fraction_vs_luminosity}. When
  $1152^3$ runs were performed the $L_C/L_\mrm{tot}$ fractions
agree with the corresponding $768^3$ runs. Because we start our analysis when a
quasi-stationary state has been reached (with the exception of the unstable run
I11), $L_\mrm{C}$ is close to $q X_\mrm{12}\dot{M}_\mrm{e}$, where $q$ is the
energy released per unit mass of $^{12}$C burnt and $X_{12} = 0.13$ the mass
fraction of $^{12}$C in the entrained material (see
Sect.~\ref{sect:3dsimsetup}). We can then write $L_\mrm{C}/L_\mrm{tot} = q
X_\mrm{12}\dot{M}_\mrm{e}/L_\mrm{tot}$ with $\dot{M}_\mrm{e}/L_\mrm{tot} = 1.36
\times 10^{-17}$\,M$_\odot$\,s$^{-1}$\,L$_\odot^{-1}$ for all runs that follow
the linear scaling relation shown in
Fig.~\ref{fig:entrainment_rate_vs_luminosity}. The \reactCC{} reaction as
implemented in \netone{} has $q_\mrm{CC1} = 0.390$\,MeV\,amu$^{-1}$ (see
Sect.~\ref{sect:reaction_network}), which implies $L_\mrm{C}/L_\mrm{tot} =
34.5\%$. This fraction is close to the measured values for runs I13 ($33.3\%$),
I14 ($31.4\%$), and I5 ($30.5\%$), which closely follow the entrainment rate
scaling. Runs with $L_\mrm{tot} \lesssim 3 \times 10^{11}$\,L$_\odot$ fall below
the scaling relation in Fig.~\ref{fig:entrainment_rate_vs_luminosity}, which
decreases $\dot{M}_\mrm{e}/L_\mrm{tot}$ and, consequently, also
$L_\mrm{C}/L_\mrm{tot}$ for these runs.

As implemented in \nettwo{}, the yields of \reactCC{} and \reactCO{} reactions
are $q_\mrm{CC2} = 0.133$\,MeV\,g$^{-1}$ and $q_\mrm{CO} =
0.438$\,MeV\,g$^{-1}$. The luminosities $L_\mrm{CC}$ and $L_{CO}$ due to the two
reactions are almost equal in run I17 ($f_\mrm{OO} = 67.5$). The average yield,
$q = 0.199$\,MeV\,amu$^{-1}$, is almost $2\times$ lower than that in the
otherwise similar run I13 ($f_\mrm{OO} = 67.5$, \netone{}), which explains the
low value of $L_C/L_\mrm{tot}$ in I17 as compared with I13 (see
Fig.~\ref{fig:L_C_fraction_vs_luminosity}). Although C burning in run I16
($f_\mrm{OO} = 2.7$, \nettwo{}) is dominated by the \reactCO{} reaction, the
average yield $q = 0.358$\,MeV\,amu$^{-1}$ is comparable with $q_\mrm{CC1} =
0.390$\,MeV\,amu$^{-1}$ of \reactCC{} (\netone{}) used in the otherwise similar
runs I4 and I15 ($f_\mrm{OO} = 2.7$).
Figure~\ref{fig:L_C_fraction_vs_luminosity} confirms that $L_C/L_\mrm{tot}$
reaches comparable values in runs I4, I15, and I16 ($L_\mrm{tot} \sim 1.5 \times
10^{11}$\,L$_\odot$).

An interesting feature of all of the I-series runs with $L_\mrm{tot} \gtrsim
3\times 10^{12}$\,L$_\odot$ is that their luminosity significantly increases in
time.  This is due to the gradual heating up of the convection zone (by ${\sim}
10\%$) combined with the nuclear reactions' high temperature sensitivity:
quantified as $\partial \log \varepsilon / \partial \log T$, the typical
sensitivities at the bottom of the convection zone are $32$ for \reactOO{}, $19$
for \reactCC{}, and 25 for \reactCO{}.

\subsection{Velocity field}
\label{sect:velocity_field}

\afterpage{
\begin{figure}
\includegraphics[width=\linewidth]{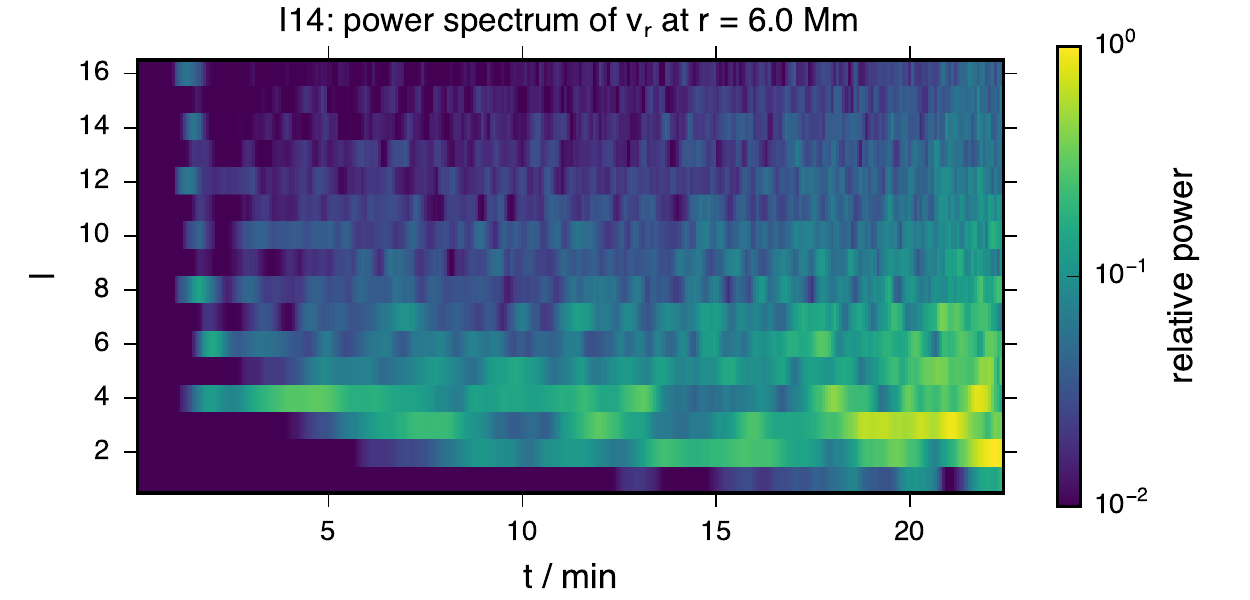}
\caption{Spherical harmonic power spectrum of the radial velocity component
$v_\mrm{r}$ at $r = 6$\,Mm in run I14 ($f_\mrm{OO} = 67.5$) shown as a function
of time.}
\label{fig:vr_spectrum_6.0_I14}
\end{figure}
\begin{figure}
\includegraphics[width=\linewidth]{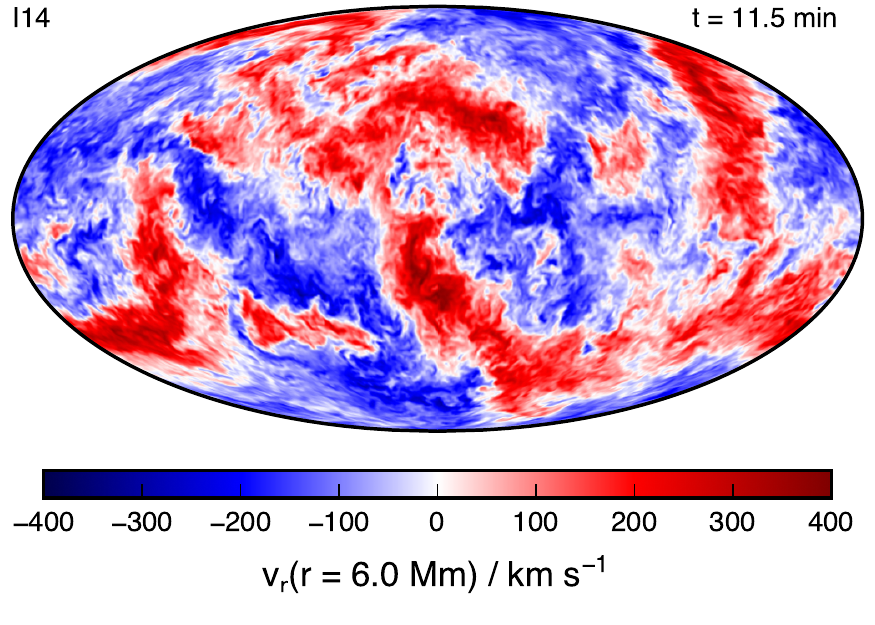}
\caption{Typical distribution of the radial velocity $v_\mrm{r}$ at $r = 6$\,Mm.
Run I14 ($f_\mrm{OO} = 67.5$) is shown in the equal-area Mollweide projection at
$t = 11.5$\,min. The rendering is based on a full 3D data cube downsampled from
the run's resolution $1152^3$ to $288^3$.}
\label{fig:vr_6.0_I14}
\end{figure}
\begin{figure}
\includegraphics[width=\linewidth]{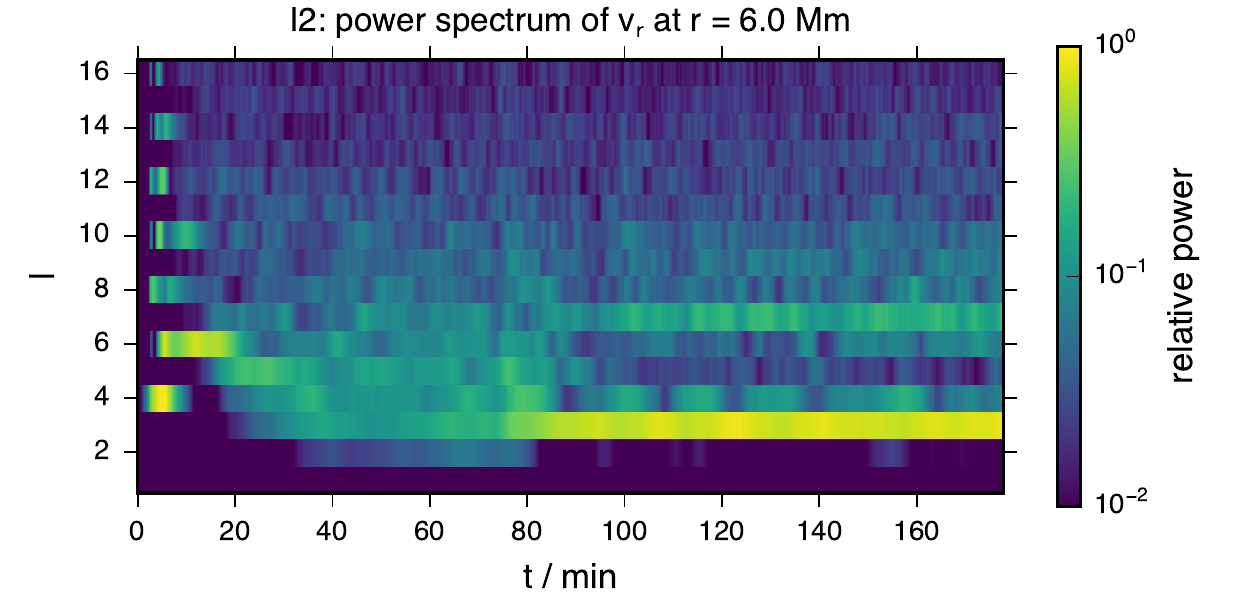}
\caption{As Fig.~\ref{fig:vr_spectrum_6.0_I14} but run I2 ($f_\mrm{OO} = 67.5$)
is shown.}
\label{fig:vr_spectrum_6.0_I2}
\end{figure}
\begin{figure}
\includegraphics[width=\linewidth]{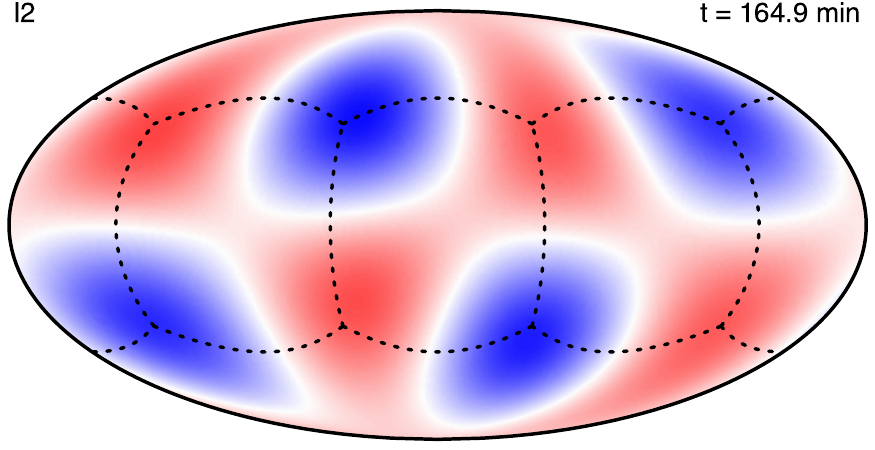}\\[0.25cm]
\includegraphics[width=\linewidth]{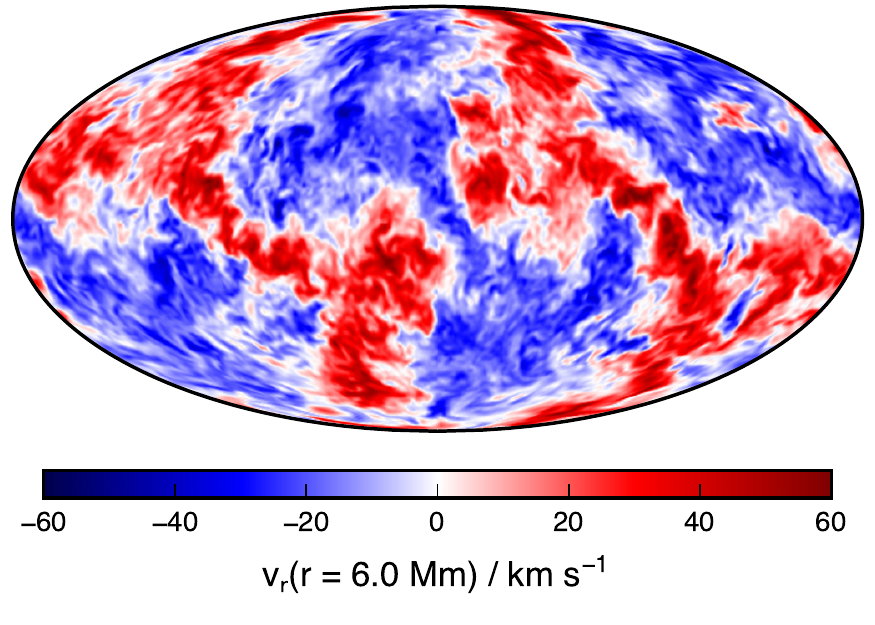}
\caption{Distribution of the radial velocity $v_\mrm{r}$ at $r = 6$\,Mm and $t =
164.9$\,min in run I2 ($f_\mrm{OO} = 67.5$, see also
Fig.~\ref{fig:vr_spectrum_6.0_I2}). The top panel only shows spherical harmonic
components of degree $l \leq 4$ based on averages in 80 triangular patches
covering the sphere (``buckets'', see Sect.~\ref{sect:ppmstar}). This pattern,
close to the spherical harmonic mode of degree $l = 3$ and order $m = -2$ and
clearly aligned with the corners of the simulation cube (dashed lines), occurs
when the driving of convection is very weak.\protect\footnotemark~Smaller scales
in the velocity field can be seen in the bottom panel, which is based on a full
3D data cube downsampled from the run's resolution $768^3$ to $192^3$.  The
equal-area Mollweide projection is used in both maps.}
\label{fig:vr_6.0_I2}
\end{figure}
\footnotetext{See also an animated version of this map and of similar maps
from other runs at the URL given in Sect.~\ref{sect:summary}.}
}

To construct the initial condition, we map a 1D hydrostatic stratification onto
a 3D Cartesian grid and set all velocity components to zero. The convective flow
is driven by the heat released from O burning at the bottom of the shell. An
initial transient flow carrying an imprint of the computational grid
disintegrates rapidly and the flow becomes fully turbulent after a few
convective overturns as described in detail by \citet{woodward15} and
\citet{jones17}. This effect can also be seen in the radial velocity spectra
shown in Figs.~\ref{fig:vr_spectrum_6.0_I14} and \ref{fig:vr_spectrum_6.0_I2},
in which an initially regular pattern of modes quickly disappears and is
replaced by a turbulent spectrum.

It is difficult to quantify the degree of turbulence in our simulations in terms
of a Reynolds number, because the PPM advection scheme does not include any
explicit viscous term and its implicit viscosity depends strongly on the ratio
$\lambda/\Delta x$ of the wavelength $\lambda$ of interest in the velocity field
and the grid cell width $\Delta x$.  \citet{porter_woodward94} measure the
dissipation properties of PPM and derive the Reynolds number
\begin{equation}
\mrm{Re_{max}} = \frac{4\pi^2}{A_s} \left( \frac{\lambda}{\Delta x}
\right)^3
\label{eq:Re_max}
\end{equation}

associated with structures on the wavelength $\lambda$, where the coefficient
$A_s$ depends on the Courant number \mbox{$\mrm{Ca} = v\Delta t/\Delta x$}. For
run I4 ($f_\mrm{OO} = 2.7$) we have a typical rms convective velocity $v =
40$\,km\,s$^{-1}$ and time step $\Delta t = 3.8 \times 10^{-3}$\,s, which gives
$\mrm{Ca} = 6.0 \times 10^{-3}$ for the advection of a typical fluid parcel. We
estimate $A_s = 47$ for this value of $\mrm{Ca}$ based on Table~1 of
\citet{porter_woodward94}. The relevant length scale $\lambda$ in our O shell is
the depth of the shell, which is $4.0$\,Mm or $157$ computational cell widths on
the $768^3$ grid. Equation~\ref{eq:Re_max} then gives $\mrm{Re_{max}} = 3.2
\times 10^6$. For the $1536^3$ grid (run D2), the Reynolds number would be
$(1536/768)^3 = 8$ times larger: $\mrm{Re_{max}} = 2.6 \times 10^7$. These large
values, although still many orders of magnitude smaller than what one would
expect for the stellar interior, reflect the low effective viscosity of the PPM
scheme on the largest flow scales. \citet{sytine00} show that the spectrum of 3D
PPM turbulence is almost indistinguishable from the expected Kolmogorov scaling
for structures with a wavelength of 32 grid cells or more, which is $1/5$ of the
depth of our convective shell on the $768^3$ grid. Strong dissipation sets in
for wavelengths smaller than about 6 grid cells or $4\%$ of the shell's depth.

Figure~\ref{fig:vr_6.0_I14} shows that the velocity field is dominated by
large-scale updraughts and downdraughts with complex small-scale turbulent
structure in the state of quasi-stationary convection. The velocity distribution
changes randomly in time and no particular direction in space is preferred, as
expected of convection in the absence of rotation and magnetic field. The
large-scale flow pattern remains essentially the same when grid resolution is
increased from $768^3$ to $1152^3$.

Several low-order modes, including an $l = 2$ mode, make approximately equal
contributions to the velocity field in most of our intermediate- and
high-luminosity runs, see runs I5 ($f_\mrm{OO} = 13.5$, \netone{}) and I13
($f_\mrm{OO} = 67.5$, \netone{}) in Fig.~\ref{fig:vr_spectra_6.0_appx} and I14
($f_\mrm{OO} = 67.5$, \netone{}) in Fig.~\ref{fig:vr_spectrum_6.0_I14}. The
high-luminosity run I17 ($f_\mrm{OO} = 67.5$, \nettwo{}) is an exception
discussed below. A strong dipolar component ($l = 1$), which one would not
expect in stationary convection in a shell with an inner radius as large as one
half of its outer radius, only appears towards the end of run I11, which shows a
very asymmetric instability (see Sect.~\ref{sect:stability}) reminiscent of the
GOSH phenomenon \citep{herwig14}. A contribution from an $l = 1$ mode is present
in the high-luminosity runs I13 and I14, but it is small and it only develops
when a large portion of the upper stable layer has been entrained and the upper
boundary of the convection zone approaches the outer boundary condition.

We observe a weak but noteworthy effect in low-luminosity runs I2 ($f_\mrm{OO} =
1$, \netone{}), I4 ($f_\mrm{OO} = 2.7$, \netone{}), I12 ($f_\mrm{OO} = 2.7$, no
C burning), I15 ($f_\mrm{OO} = 2.7$, \netone{}), and I16 ($f_\mrm{OO} = 2.7$,
\nettwo{}). The large-scale flows keep changing their distribution for
${\sim}10$\,--\,$15$ overturning time scales, after which they start to converge
towards the $l = 3$ mode of order $m = -2$ with large-scale updraughts and
downdraughts aligned with the diagonals of the simulation cube. The flow remains
highly turbulent and it is sometimes difficult to recognise this pattern by eye,
but it can nevertheless be identified using spherical harmonic analysis
(Fig.~\ref{fig:vr_spectrum_6.0_I2}) and visualised by applying a low-pass filter
to the velocity distribution (Fig.~\ref{fig:vr_6.0_I2}). It is suspicious that
the grid aligned pattern first appears at $t \approx 80$\,min in I2, because we
turned on C burning in that run at $t = 74$\,min, see
Sect.~\ref{sect:simulations}. However, we see the same pattern in runs I4, I15,
and I16 that had C burning on from $t = 0$ and even in run I12 that had C
burning turned off throughout. It is possible that the geometry of our shell
slightly prefers convection cells about as large as those of an $l = 3$ mode
and, if the convection is weakly driven, these cells align themselves with the
$l = 3$ mode that is slightly preferred by our Cartesian grid. We are currently
investigating such subtle effects that occur in slow flows using a new version
of the \ppmstar{} code, which solves for fluctuations around a base state. Our
preliminary results indicate that this formulation offers advantages for
problems involving weakly driven convection.

Since the grid-alignment occurs in the phase of quasi-stationary convection, one
might suspect that this effect could somehow influence the entrainment rate,
which we measure in this phase, and it could possibly explain why the
entrainment rates are lower than what our scaling law predicts at low
luminosities, see Sect.~\ref{sect:entrainment_rate} and
Fig.~\ref{fig:entrainment_rate_vs_luminosity}. However, we find that the
entrainment rate slightly increases in the early evolution towards the
quasi-stationary state in all of the relevant runs, see run I2 ($f_\mrm{OO} =
1$, \netone{}) in Fig.~\ref{fig:entrainment_rate}, I4 ($f_\mrm{OO} = 2.7$,
\netone{}) in Fig.~\ref{fig:entrainment_rate_appx}, and runs I15 ($f_\mrm{OO} =
2.7$, \netone{}) and I16 ($f_\mrm{OO} = 2.7$, \nettwo{}) in
Fig.~\ref{fig:entrainment_rate_appx2}. Therefore, if the grid-alignment effect
has any influence on the entrainment rate, it most likely causes a slight
increase and not a decrease.

Our choice of the C-burning network (\netone{} vs \nettwo{}) does not influence
the velocity distribution at the low luminosity of runs I4 ($f_\mrm{OO} = 2.7$,
\netone{}) and I16 ($f_\mrm{OO} = 2.7$, \nettwo{}). Velocity spectra from these
two runs are essentially the same, see Fig.~\ref{fig:vr_spectra_6.0_appx}. This
is not the case for the pair of high-luminosity runs I13 ($f_\mrm{OO} = 67.5$,
\netone{}) and I17 ($f_\mrm{OO} = 67.5$, \nettwo{}). I13 switches between modes
$l = 2, 3, 4$ at random whereas I17 is dominated by an $l = 3$ mode throughout
most of it evolution. Convection cells in this pattern, however, keep moving
around the sphere unlike those in the low-luminosity runs mentioned above. We do
not know the cause of this effect.

\begin{figure}
\includegraphics[width=\linewidth]{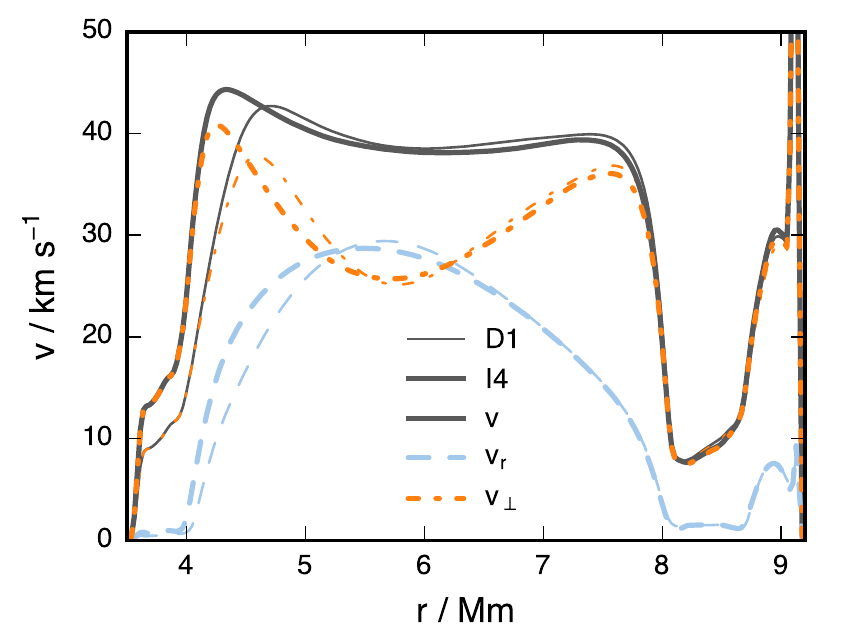}
\caption{Total ($v$), radial ($v_\mrm{r}$), and tangential ($v_\perp$) rms
velocities in run I4 ($f_\mrm{OO} = 2.7$) as compared with run D1 of J17.}
\label{fig:velocity_profiles_I4_vs_D1}
\end{figure}

\begin{figure}
\includegraphics[width=\linewidth]{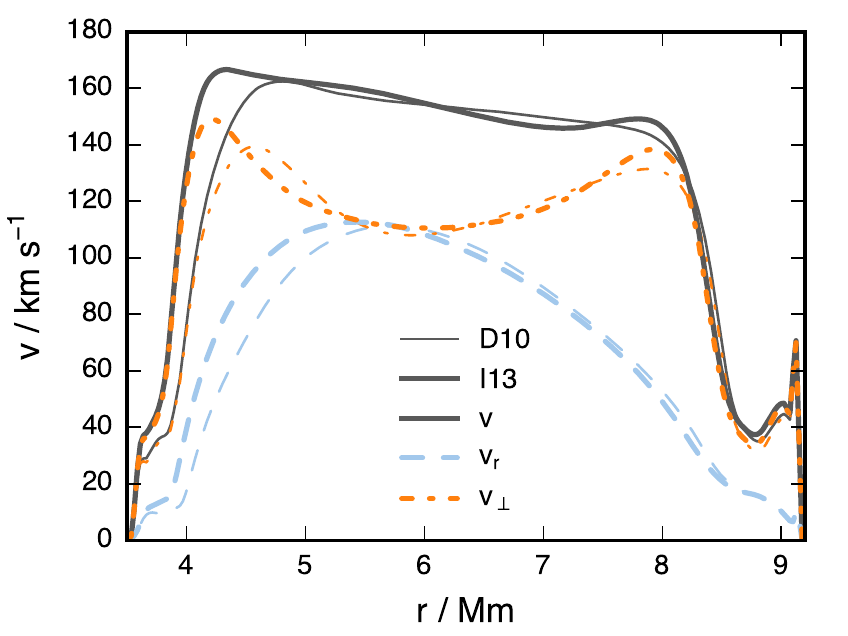}
\caption{As Fig.~\ref{fig:velocity_profiles_I4_vs_D1}, but runs I13 ($f_\mrm{OO}
= 67.5$) and D10 are shown.}
\label{fig:velocity_profiles_I13_vs_D10}
\end{figure}

The fact that we use a realistic O-burning prescription and include the burning
of the entrained material in our simulations as opposed to the simulations
presented by J17 makes surprisingly little difference for time-averaged velocity
profiles. Figure~\ref{fig:velocity_profiles_I4_vs_D1} compares the
low-luminosity run I4 ($f_\mrm{OO} = 2.7$) with the similar run D1 of J17. The
only difference can be seen at the transition from the lower stable layer into
the convection zone, which is located deeper and is  more abrupt in I4 than in
D1. This is qualitatively compatible with the heating rate distribution's being
more concentrated to the bottom of the convection zone in the I series of runs,
see Figs.~\ref{fig:heating_distribution} and \ref{fig:energy_sources_I2_I13}. We
can see the same effect in the comparison between the high-luminosity runs D10
and I13 ($f_\mrm{OO} = 67.5$) in Fig.~\ref{fig:velocity_profiles_I13_vs_D10}.

\begin{figure}
\includegraphics[width=\linewidth]{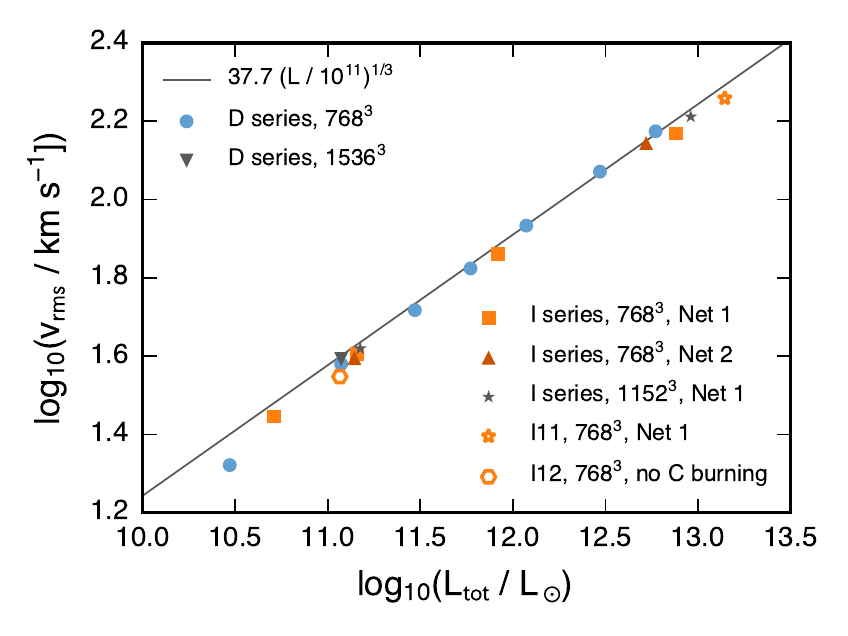}
\caption{Dependence of the rms convective velocity $v_\mathrm{rms}$ on the total
luminosity $L_\mrm{tot}$ integrated over the convection zone. The
expected scaling law $v_\mathrm{rms} \propto L_\mathrm{tot}^{1/3}$ is shown for
comparison. The last $2.2$\,min of run I11 are used when the instability
described in Sect.~\ref{sect:stability} has fully developed.}
\label{fig:rms_velocity_vs_luminosity}
\end{figure}

Convective velocity is expected to scale with $L_\mrm{tot}^{1/3}$ on theoretical
grounds \citep[][J17]{biermann32, porter_woodward00, muller_janka15}, where
$L_\mrm{tot}$ is the total luminosity driving convection. We calculate the rms
velocity $v_\mrm{rms} = (2 E_\mrm{k,cz} / M_\mrm{cz})^{1/2}$, where
$E_\mrm{k,cz}$ and $M_\mrm{cz}$ are the total kinetic energy and mass of the
convection zone, respectively. We average $v_\mrm{rms}$ over the time interval
during which the entrainment rate is measured, see
Figs.~\ref{fig:entrainment_rate}, \ref{fig:entrainment_rate_appx}, and
\ref{fig:entrainment_rate_appx2}. Figure~\ref{fig:rms_velocity_vs_luminosity}
shows that all runs, including the unstable run I11 (see
Sect.~\ref{sect:stability} for details), closely follow the expected scaling law
and only the least luminous run D23 starts to deviate from it (by ${\sim}16\%$).

\subsection{Fluctuations}
\label{sect:fluctuations}

The 3D, time-dependent nature of the flow gives rise to a whole spectrum of
fluctuations. Their magnitude depends on the total luminosity, which both
determines and depends on the entrainment rate. To quantify these fluctuations,
we have used averages measured in 80 different directions (`` buckets'', see
Sect.~\ref{sect:ppmstar} for details).

\begin{figure}
\includegraphics[width=\linewidth]{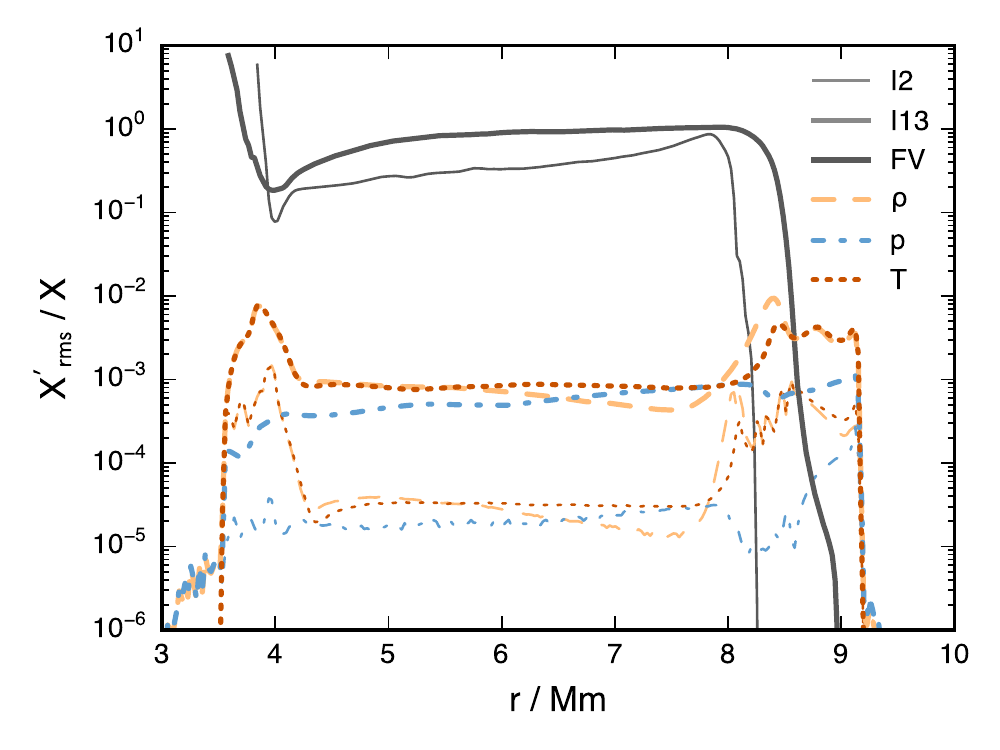}
\caption{Relative large-scale rms fluctuations of the fractional volume FV of
fluid \cldfluid{}, density $\rho$, pressure $p$, and temperature $T$ as
determined from 80 radial \emph{buckets} that together cover the full solid
angle (see Sect.~\ref{sect:ppmstar} for details). Runs I2 ($f_\mathrm{OO} =
1$, \netone{}) and I13 ($f_\mathrm{OO} = 67.5$, \netone{}) are shown at $t =
134.9$\,min and $t = 9.32$\,min, respectively, when I13 is ${\sim} 150$ times
more luminous than I2.}
\label{fig:fluctuations_I2_I13}
\end{figure}

Figure~\ref{fig:fluctuations_I2_I13} compares the low-luminosity run I2
($f_\mrm{OO} = 1$) with the high-luminosity run I13 ($f_\mrm{OO} = 67.5$) in
terms of the relative rms fluctuations $X^\prime_\mrm{rms}(r) / X(r)$, where $X$
is the quantity of interest, $X(r)$ the spherical average of $X$ at the radius
of $r$, and
\begin{equation}
X^\prime_\mrm{rms}(r) = \sqrt{\frac{1}{N}\sum_{i=1}^{N} \left[X_i(r) -
X(r)\right]^2},
\label{eq:rms_fluctuations}
\end{equation}
defines the absolute rms fluctuations, where $N = 80$ is the total number of
buckets. Fluctuations in the density $\rho$, pressure $p$, and temperature $T$
are the same within a factor of a few in the convection zone and they increase
with increasing luminosity. Pressure fluctuations are much smaller than density
fluctuations in the stable layers as expected for buoyancy-driven internal
waves.

Density fluctuations in the convection zone are close to $\mrm{Ma}^2$ in both I2
and I13 (see Table~\ref{tab:runs} for the $\mrm{Ma}$ values). The contrast in
magnitude between the density and temperature fluctuations at the convective
boundaries and those in the convection zone becomes larger with decreasing
luminosity and Mach number. \citet{meakin_arnett07} estimate the magnitude of
density fluctuations in a convection zone and at its boundaries as $\rho^\prime
/ \rho \sim \mrm{Ma}^2 + v_\mrm{conv}\,N / g$, where $\mrm{Ma}$ is the
convective Mach number, $v_\mrm{conv}$ the convective velocity, $N$ the local
Brunt-V\"ais\"al\"a frequency, and $g$ the gravitational acceleration. The
$v_\mrm{conv}\,N / g$ term is proportional to $\mrm{Ma}$, so the relative
magnitude of density fluctuations in the stable layers as compared to those in
the convection zone is inversely proportional to $\mrm{Ma}$ (c.f.\
Fig.~\ref{fig:fluctuations_I2_I13}). Both convective boundaries in our
simulations are characterised by approximately the same ratio $g / N \approx
4500$\,km\,s$^{-1}$. Substituting $v_\mrm{rms}$ from Table~\ref{tab:runs} for
$v_\mrm{conv}$, we get that the contribution to $\rho^\prime / \rho$ from the
term $v_\mrm{conv}\,N / g = v_\mrm{conv} / (g / N)$ in the expression above is
$\approx 6 \times 10^{-3}$ in I2 and $\approx 3 \times 10^{-2}$ in I13. These
values are significantly larger than our measurements shown in
Fig.~\ref{fig:fluctuations_I2_I13}, which implies that the order-of-magnitude
expression $v_\mrm{conv}\,N / g$ overestimates the actual fluctuations on the
bucket-to-bucket scale.\footnote{We have sub-bucket-scale rms information for
the fractional volume only, but it is not included in
Fig.~\ref{fig:fluctuations_I2_I13} for consistency with the other variables
shown.}

\begin{figure}
\includegraphics[width=\linewidth]{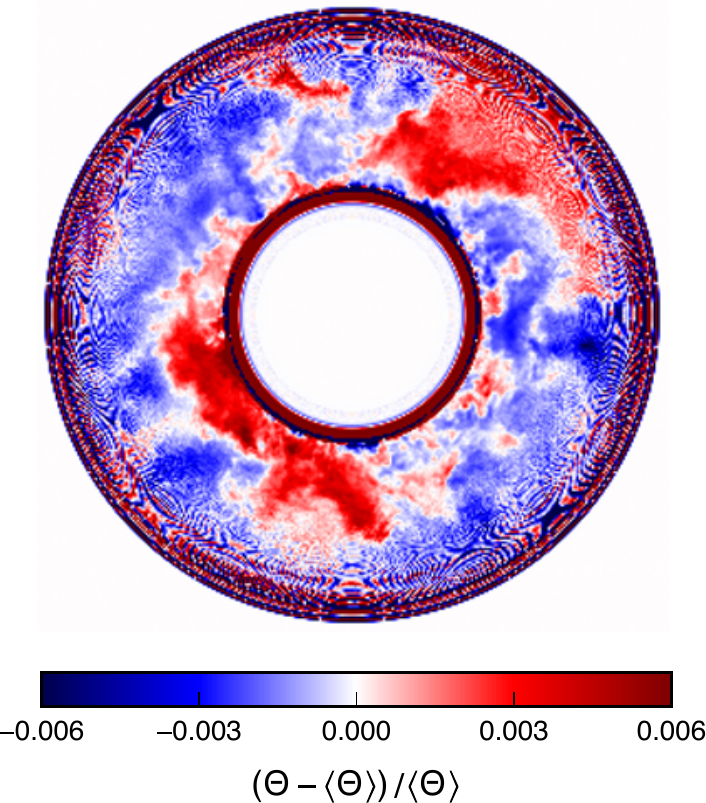}
\caption{Relative fluctuations in the corrected temperature $\Theta$ (see
Eq.~\ref{eq:T9_drop_2} and Sect.~\ref{sect:reaction_network}) with respect to
the spherical average $\langle \Theta \rangle$ at $t = 21.3$\,min in run I14
($f_\mrm{OO} = 2.7$). The fluctuations are averaged over an XZ slice $1$\,Mm
thick ($5\%$ of the simulation box width). The rendering is based on data
downsampled from $1152^3$ to $288^3$ that were further subject to a lossy
compression algorithm (see Sect.~\ref{sect:ppmstar} for details). Numerical
noise from the compression step dominates the outer portion of the rendering.}
\label{fig:T9_fluctuations_I14}
\end{figure}
We plot the uncorrected temperature $T$ in Fig.~\ref{fig:fluctuations_I2_I13},
although nuclear reaction rates are computed using the corrected temperature
$\Theta$ (Eqs.~\ref{eq:T9_drop_1} and \ref{eq:T9_drop_2}).
Figure~\ref{fig:T9_fluctuations_I14} shows relative fluctuations in $\Theta$
close to the end of run I14 (one of the most luminous, $f_\mrm{OO} = 67.5$).
They are smaller than $1\%$, which limits their possible influence on the
C-burning rates to $\lesssim 20\%$ even at the upper end of the luminosity range
considered (see Sect.~\ref{sect:luminosity_evolution} for the reactions'
temperature sensitivities). For this reason, we did not include them in
Fig.~\ref{fig:renderings_I2_I13}.

\begin{figure}
\includegraphics[width=\linewidth]{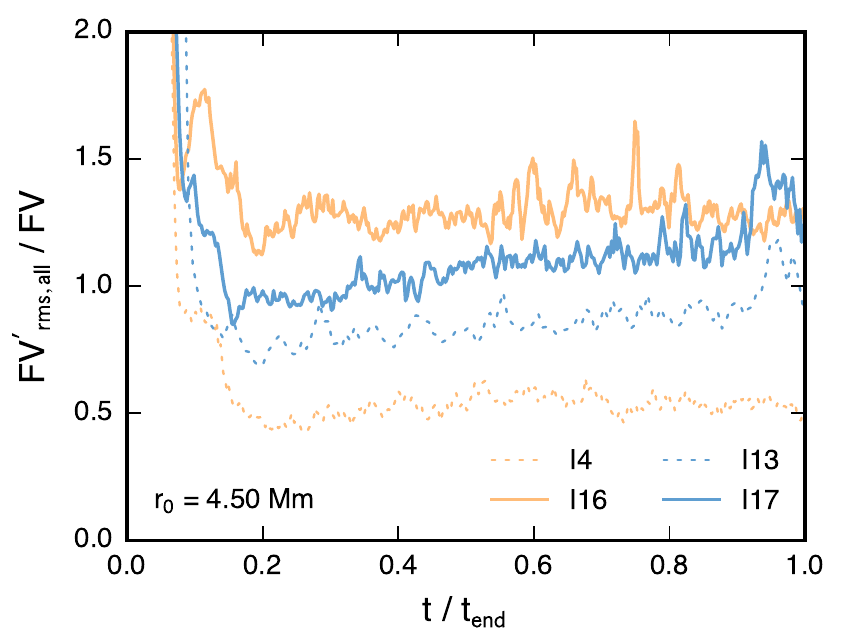}
\caption{Time evolution of the relative rms fluctuations in the fractional
volume FV of fluid \cldfluid{} at the radius of $r_0 = 4.5$\,Mm in runs I4
($f_\mathrm{OO} = 2.7$, \netone{}), I13 ($f_\mathrm{OO} = 67.5$, \netone{}), I16
($f_\mathrm{OO} = 2.7$, \nettwo{}), and I17 ($f_\mathrm{OO} = 67.5$, \nettwo{}).
All spatial scales are considered here as opposed to
Fig.~\ref{fig:fluctuations_I2_I13}. Simulation time on the abscissa is
normalised for clarity.}
\label{fig:flow_inhomogeneity}
\end{figure}

Fluctuations in the fractional volume $\mrm{FV}$ of fluid \cldfluid{} are large
in all of our runs, including the least luminous ones (see
Fig.~\ref{fig:fluctuations_I2_I13} and also Fig.~\ref{fig:renderings_I2_I13}).
They slightly increase with increasing luminosity, although much less than the
fluctuations in $\rho$, $p$, and $T$ do, and they create large-scale asymmetries
in the energy generation rate as shown in Fig.~\ref{fig:renderings_I2_I13}. The
$\mrm{FV}$ fluctuations in runs I16 ($f_\mrm{OO} = 2.7$) and I17 ($f_\mrm{OO} =
67.5$), which use the C-burning network \nettwo{}, are larger than those in a
similar pair of runs I4 ($f_\mrm{OO} = 2.7$) and I13 ($f_\mrm{OO} = 67.5$), in
which \netone{} is used, see Fig.~\ref{fig:flow_inhomogeneity}. This is likely a
consequence of the fact that a certain concentration $X_{12}$ of $^{12}$C has to
first build up in the convection zone to make the time scale of the \reactCC{}
reactions ($\propto X_{12}^{-1}$) considered in \netone{} short enough to
balance the rate of mass entrainment whereas the C-burning time scale is
independent of $X_{12}$ for the \reactCO{} reactions that are also included in
\nettwo{}. \reactCO{} reactions are responsible for ${\sim}50\%$ of the
C-burning luminosity in run I17 and they dominate C burning in run I16, see
Sect.~\ref{sect:luminosity_evolution}. The $\mrm{FV}$ fluctuations as quantified
in Fig.~\ref{fig:flow_inhomogeneity} are slightly larger than those in
Fig.~\ref{fig:fluctuations_I2_I13} (compare run I13 shown in both), because all
spatial scales are considered (i.e.\ the index $i$ in
Eq.~\ref{eq:rms_fluctuations} runs over computational grid cells instead of the
buckets) in Fig.~\ref{fig:flow_inhomogeneity} as opposed to
Fig.~\ref{fig:fluctuations_I2_I13}.

\subsection{Instability in a case with strong feedback as in O-C shell merger
conditions}
\label{sect:stability}

\begin{figure}
\includegraphics[width=\linewidth]{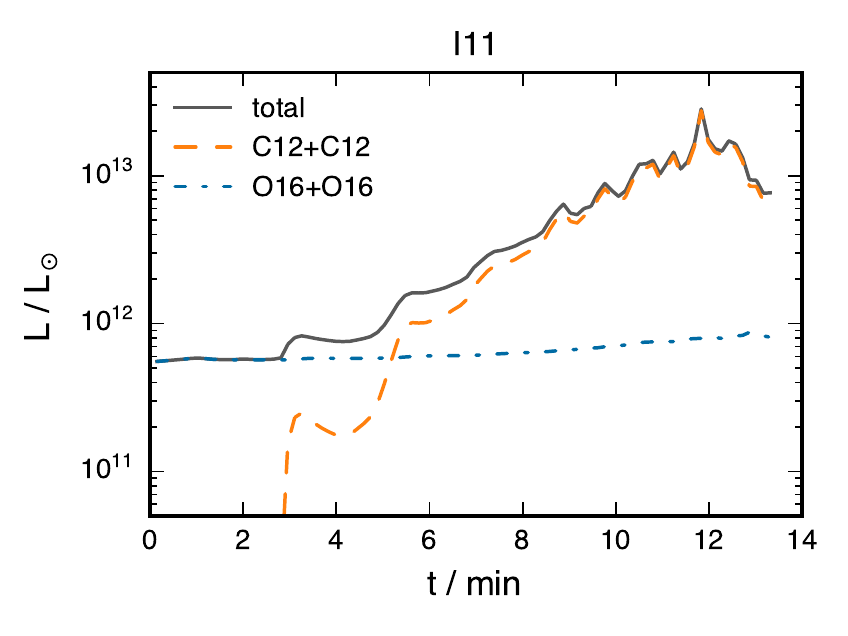}
\caption{Contributions to the total luminosity integrated over the convection
zone in run I11 ($f_\mrm{OO} = 13.5$, $f_\mrm{CC} = 10$).}
\label{fig:luminosity_I11}
\end{figure}
\begin{figure}
\includegraphics[width=\linewidth]{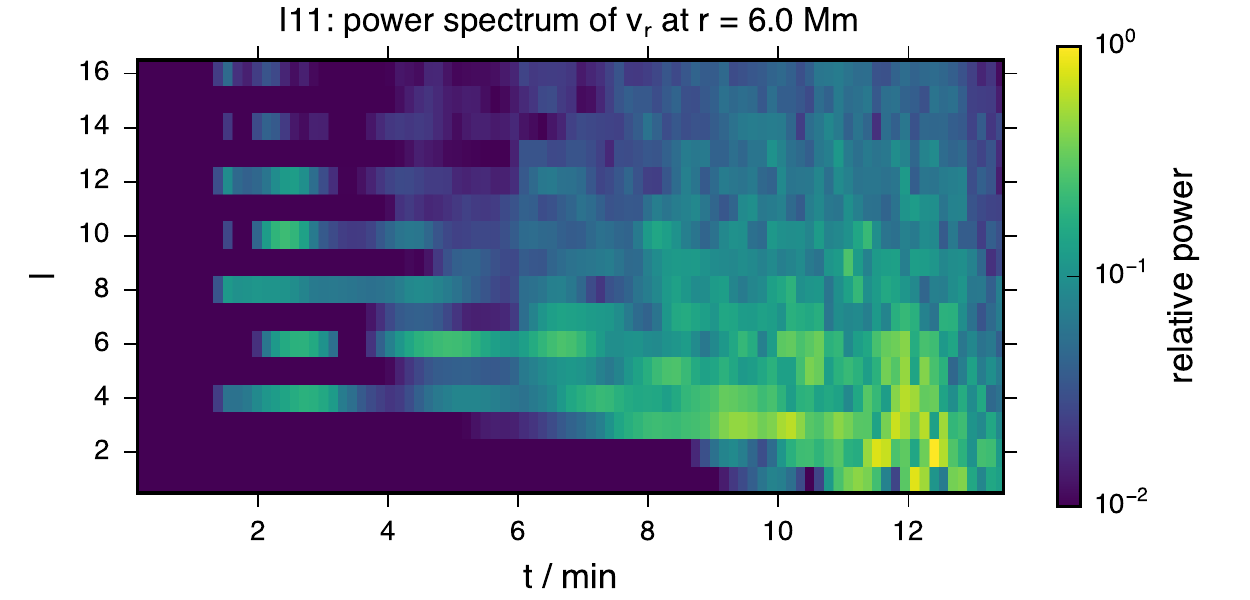}
\caption{Time evolution of the spherical harmonic power spectrum of the radial
velocity $v_\mrm{r}$ at the radius of $6$\,Mm in run I11 ($f_\mrm{OO} = 13.5$,
$f_\mrm{CC} = 10$). Spectral power is normalised by its maximum value.}
\label{fig:vr_spectrum_6.0_I11}
\end{figure}
\begin{figure}
\includegraphics[width=\linewidth]{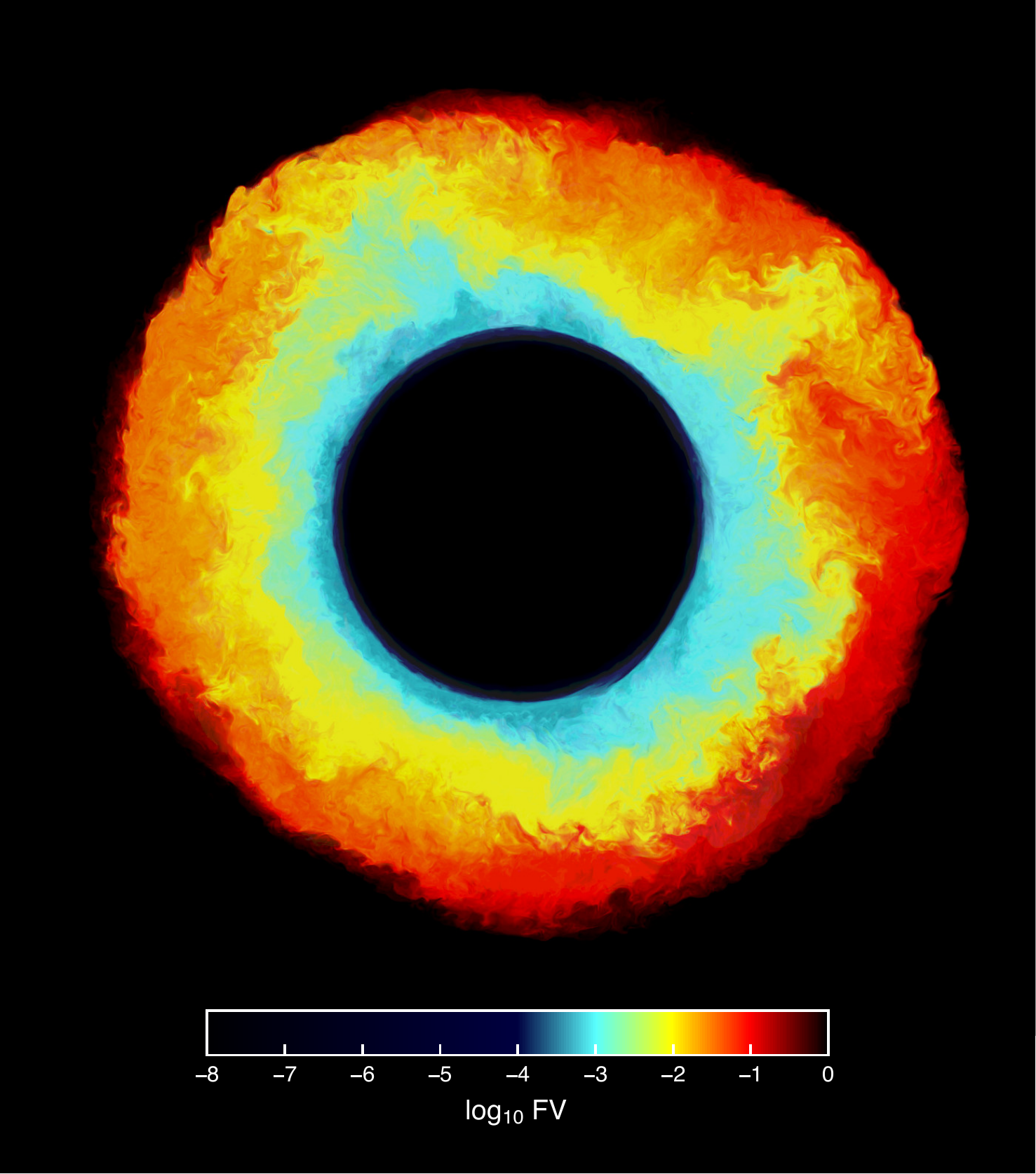}
\centering
\caption{Rendering of the fractional volume of fluid \cldfluid{} in a thin slice
through the computational box at $t = 12.6$\,min in run I11 ($f_\mrm{OO} =
13.5$, $f_\mrm{CC} = 10$).}
\label{fig:I11_FV-0085}
\end{figure}
Depending on the entropy difference between a pair of interacting O- and
C-burning convective shells, the rate of mass entrainment into the O shell and
the corresponding feedback from C-burning could be even larger than in the
simulations discussed so far. In terms of the nucleosynthetic signature of the
O-C shell merger, \citet{ritter18} show that the production of odd-Z elements
increases with increasing entrainment rate and mixing efficiency, and they
suggest that the largest enhancements could be reached in a case in which the O
and C shells merge completely on the dynamical time scale. An entrainment rate
of ${\sim}1$\,$M_\odot$\,s$^{-1}$ could be reached assuming near-sonic mass
exchange, which we consider an upper limit. To explore this regime within the
framework of the current series of runs, we have carried out a variation of the
medium-luminosity run I5 ($f_\mrm{OO} = 13.5$) which has an entrainment rate of
${\sim}10^{-5}$\,M$_\odot$\,s$^{-1}$. In order to mimic the energy feedback
encountered in a situation closer to a complete merger, we increase the energy
yield of C burning by the factor $f_\mrm{CC} = 10$ in run I11. Even taking into
account that we increased the amount of C in the ingested material by a factor
of five compared to the stellar model value (Sect.~\ref{sect:3dsimsetup}), the
resulting entrainment rate (see below), which sets the strength of the energy
feedback, stays well below the above mentioned upper limit. The degree of
asymmetric perturbations reported in this section for run I11 may therefore be
considered as a lower limit for the case of a complete O-C shell merger.

\begin{figure*}
\begin{center}
\includegraphics[width=0.9\linewidth]{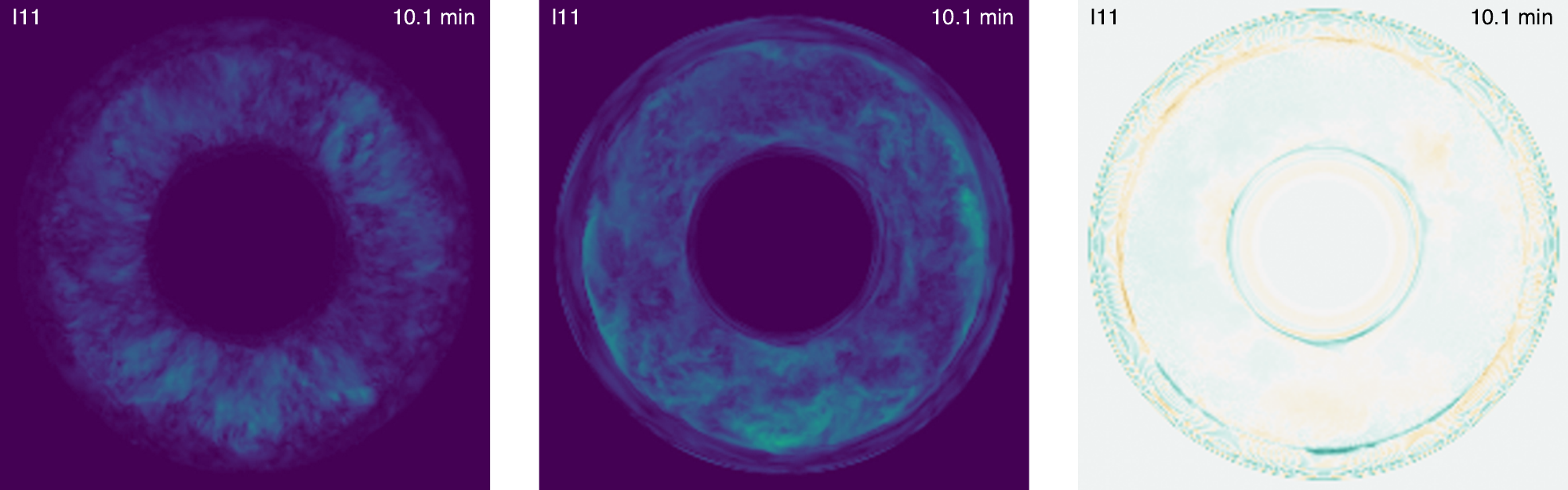}\\[0.25cm]
\includegraphics[width=0.9\linewidth]{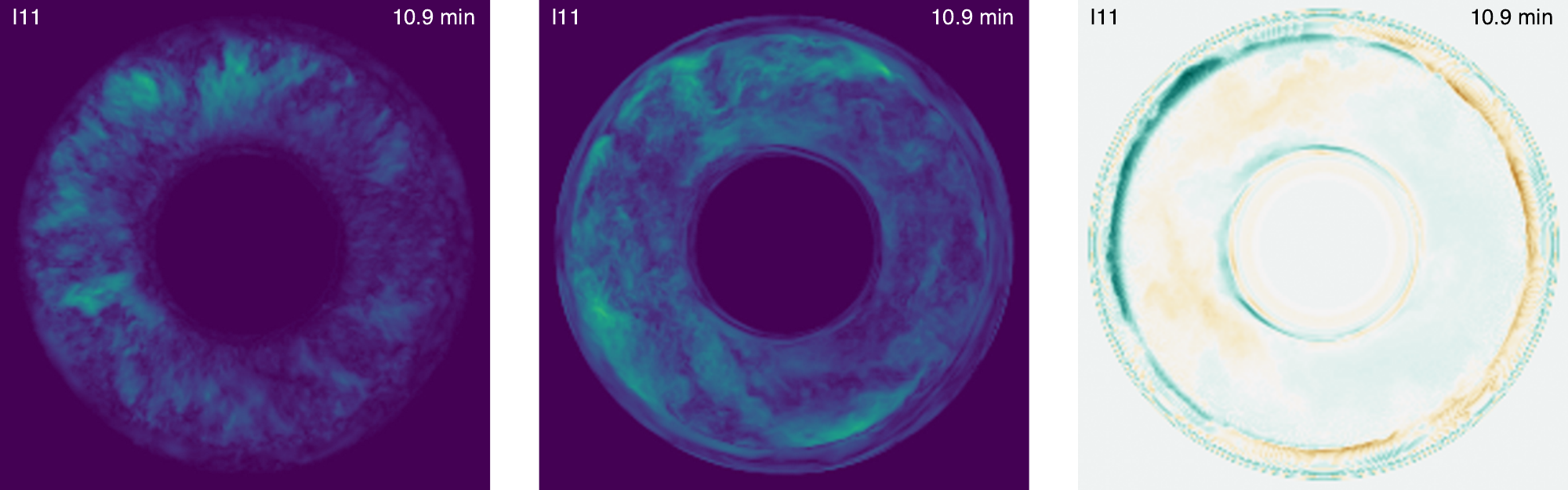}\\[0.25cm]
\includegraphics[width=0.9\linewidth]{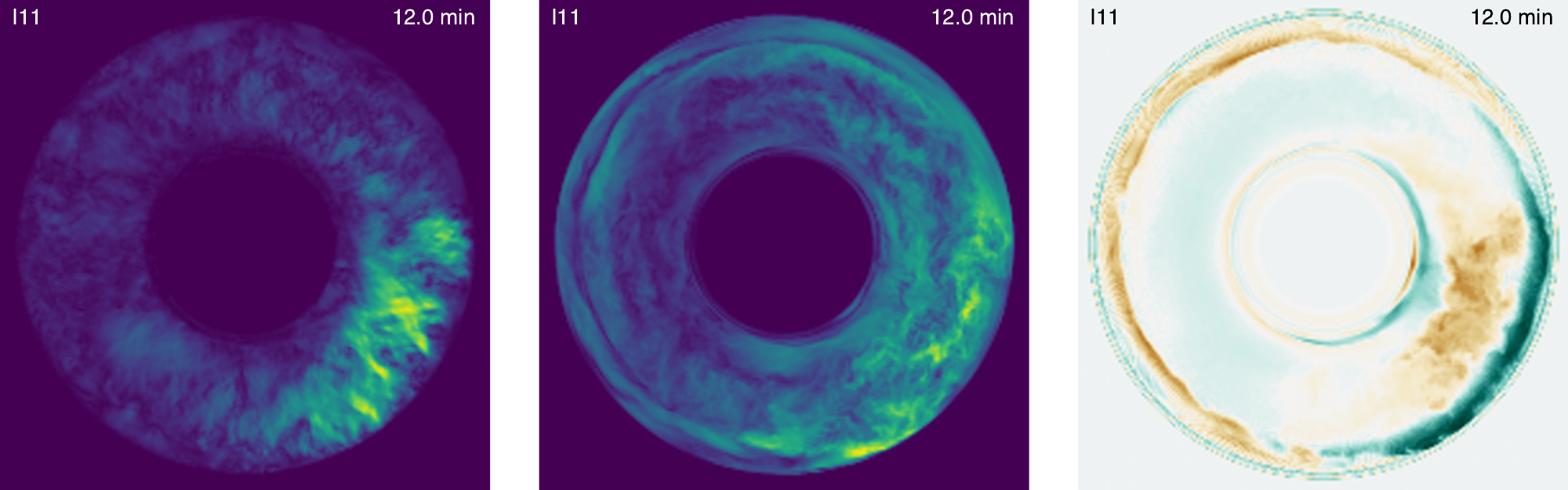}\\[0.25cm]
\includegraphics[width=0.9\linewidth]{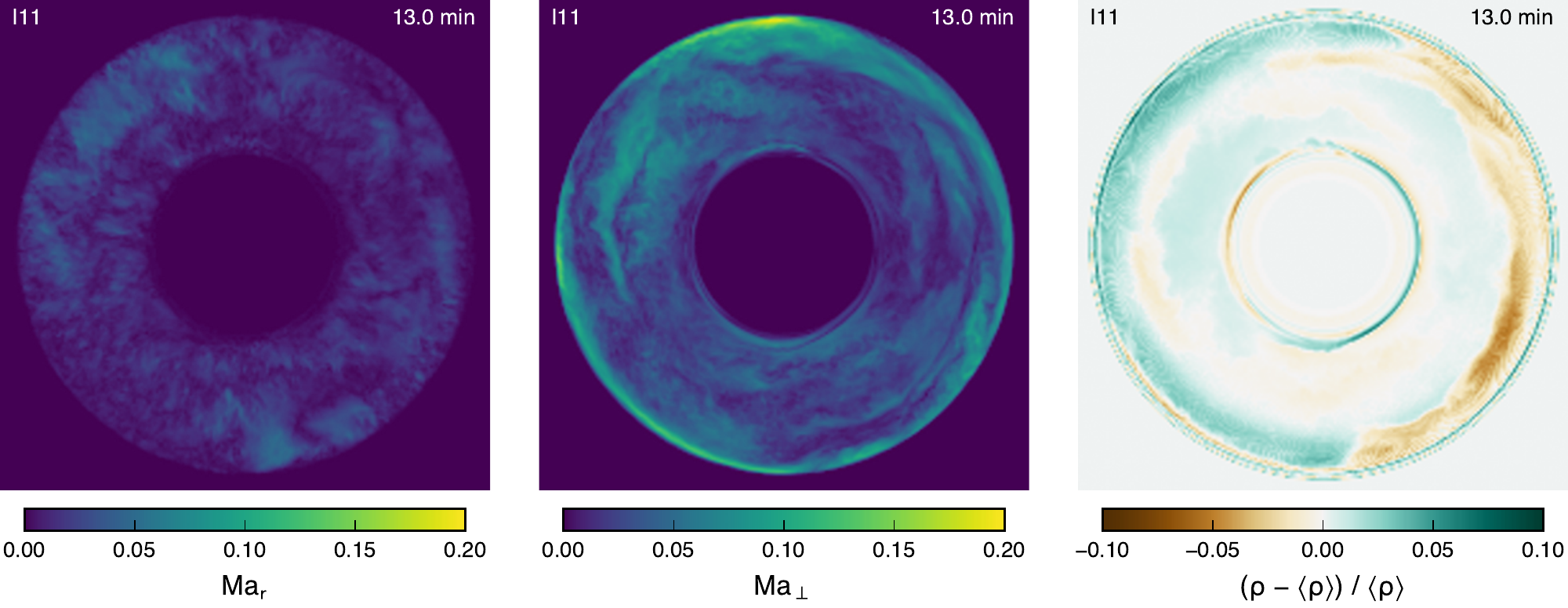}
\end{center}
\caption{Time evolution of the Mach numbers $\mrm{Ma_r}$ and
$\mrm{Ma}_\perp$ corresponding to motions in the radial and tangential
direction, respectively, and of the relative density fluctuations $(\rho -
\langle\rho\rangle) / \langle\rho\rangle$ in the last three minutes of run I11
($f_\mrm{OO} = 67.5$, $f_\mrm{CC} = 10$).}
\label{fig:Mar_Mat_rho_fluct_I11}
\end{figure*}

Figure~\ref{fig:luminosity_I11} shows that the C-burning luminosity (\reactCC{})
starts rapidly increasing soon after the onset of entrainment in I11. C burning
quickly becomes the dominant source of energy and exponential growth ensues.
Strong, large-scale oscillations akin to the GOSH phenomenon described by
\citet{herwig14} develop after $t \approx 11$\,min. In terms of spherical
harmonics, modes with $l = 1$ to $l = 4$ are dominant (see
Fig.~\ref{fig:vr_spectrum_6.0_I11}) and the upper convective boundary gets
significantly deformed (see Fig.~\ref{fig:I11_FV-0085}). The oscillations can
also be seen in Fig.~\ref{fig:Mar_Mat_rho_fluct_I11}, which shows at four points
in time the Mach numbers $\mrm{Ma_r}$ and $\mrm{Ma}_\perp$ corresponding to
motions in the radial and tangential direction, respectively, and the relative
density fluctuations $(\rho - \langle\rho\rangle) / \langle\rho\rangle$. High
values of $\mrm{Ma_r}$ and $\mrm{Ma}_\perp$, reaching ${\sim}0.3$ locally, are
concentrated to one half of the renderings at $t = 10.9$\,min and at $t =
12.0$\,min. The density fluctuations follow the same pattern. They reach values
up to ${\sim}15\%$ in some parts of the upper boundary due to the boundary's
large-scale deformation, but they also approach $5$\,--\,$10\%$ in the bulk of
the convection zone at $t = 12.0$\,min. The Courant number exceeds unity at some
point in the simulation volume at $t = 13.5$\,min and the simulation is stopped.
Figure~\ref{fig:Mar_Mat_rho_fluct_I11} also shows that although the convective
boundary is well separated from the outer spherical wall boundary condition
during the initial growth of the instability, some parts of the flow reach the
outer wall by $t = 12$\,min and are likely to be influenced by its presence at
the end of the run. This results from this run's being more violent than
anticipated.

\begin{figure}
\includegraphics[width=\linewidth]{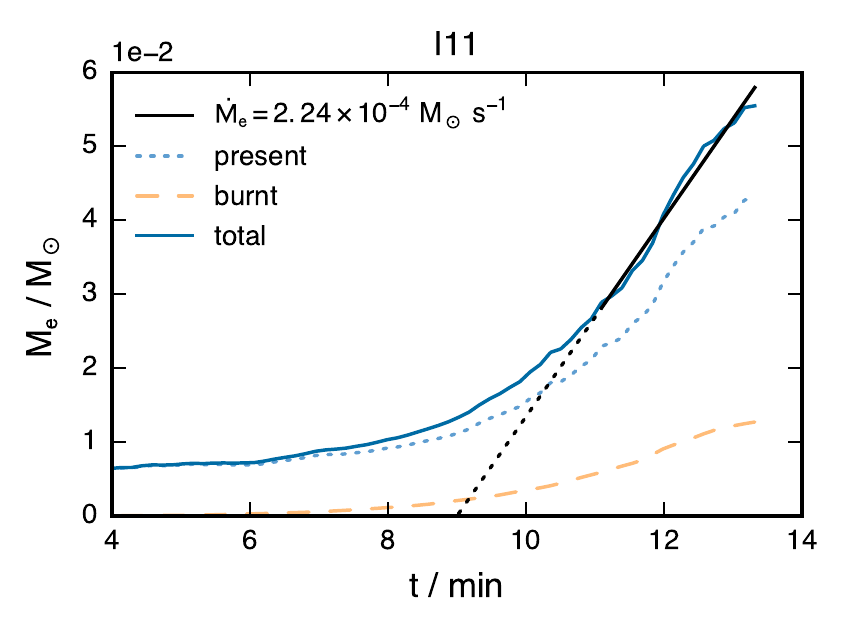}
\caption{Like Fig.~\ref{fig:entrainment_rate}, but run I11 ($f_\mrm{OO} = 67.5$,
$f_\mrm{CC} = 10$) is shown.}
\label{fig:entrainment_rate_I11}
\end{figure}

Our standard method of determining the entrainment rate is not applicable to
cases of very violent convection associated with strong motions in the upper
stable layer as mentioned in Sect.~\ref{sect:entrainment_rate}. In order to
characterise the entrainment process throughout run I11, we have slightly
modified this method. Instead of using the average velocity profile to define
the upper boundary of the convection zone, we use the bucket data (see
Sect.~\ref{sect:ppmstar}) to measure the radius of the boundary using fractional
volume profiles of fluid \cldfluid{} in 80 different directions. We define the
radius of the upper boundary in each individual bucket as the largest radius
inside that the fractional volume does not exceed $\frac{1}{2}$ and we integrate
the amount of fluid \cldfluid{} inside that radius. We have compared this method
with that described in Sect.~\ref{sect:entrainment_rate} in the time interval
from $6.2$\,min to $9.3$\,min, in which both methods are applicable.  Our
standard method yields $\dot{M}_\mrm{e} = 3.07 \times
10^{-5}$\,M$_\odot$\,s$^{-1}$ in this time interval. With the new method, we
obtain $\dot{M}_\mrm{e} = 3.61 \times 10^{-5}$\,M$_\odot$\,s$^{-1}$, i.e.\ a
value only $18\%$ larger. The new method allows us the measure the ultimate
entrainment rate achieved in the last two minutes of run I11: $\dot{M}_\mrm{e} =
2.24 \times 10^{-4}$\,M$_\odot$\,s$^{-1}$, see
Fig.~\ref{fig:entrainment_rate_I11}.

This entrainment rate is similar to the 1D nucleosynthesis run Sm4 of
\citet{ritter18} that produces only modest enhancements of odd-Z elements. It is
also similar to our high-luminosity runs I13 and I14 ($f_\mrm{OO} = 67.5$, no
C-burning enhancement), which do not show the runaway effect observed in I11.
Although mass entrainment and the subsequent nuclear burning are not spherically
symmetric in any of our 3D runs, the use of a 1D mixing prescription for
nucleosynthesis post-processing seems more justified in quasi-stationary cases
like I13 or I14 than in the unstable case I11.

Although run I11 reaches a luminosity similar to that of runs I13, I14, and I17
(compare Fig.~\ref{fig:luminosity_I11} with Figs.~\ref{fig:luminosity_evolution}
and \ref{fig:luminosity_evolution_appx}), it differs from them qualitatively.
Convection in the latter three runs is quasi-stationary and predominantly driven
by O burning with O luminosity increasing on the time scale of many convective
overturns, following the heating up of the convection zone. The rates of change
of the luminosity and entrainment rate are so rapid in I11 that this run cannot
be considered quasi-stationary. Figure~\ref{fig:entrainment_rate_I11} also shows
that $M_\mrm{e}(t)$ in I11, unlike any in other run, is dominated by the
entrained fluid's piling up in the convection zone. The total mass of fluid
\cldfluid{} entrained by the end of run I11 corresponds to only about
$\sfrac{1}{3}$ of the upper stable layer's initial mass whereas all of that
layer gets ultimately entrained in runs I13, I14, and I17. The contribution of
the temperature increase to the increase in the burning rates is rather limited
in I11 as evidenced by the small ($\lesssim 50\%$) increase in the oxygen
luminosity (see Fig.~\ref{fig:luminosity_I11}), which is a sensitive temperature
indicator.

The behaviour of run I11 can be motivated as follows. The linearity of the
dependence of the entrainment rate $\dot{M}_\mrm{e}$ on the total luminosity
$L_\mrm{tot}$ (see Fig.~\ref{fig:entrainment_rate_vs_luminosity} and
Sect.~\ref{sect:entrainment_rate}) suggests that the amount of energy needed to
entrain a constant amount of mass is constant. The scaling relation gives
$\dot{M}_\mrm{e,0} = 1.44 \times 10^{-6}$\,M$_\odot$\,s$^{-1}$ at $L_0 =
10^{11}$\,L$_\odot$, so $\Gamma = L_0 / \dot{M}_\mrm{e,0} = 1.34 \times
10^{17}$\,erg\,g$^{-1}$, which can also be expressed as $\Gamma =
0.139$\,MeV\,amu$^{-1}$. Only the mass fraction $X_\mrm{C12} = 0.13$ of $^{12}$C
in the entrained fluid represents burnable fuel, so $\Gamma_\mrm{C12} = \Gamma /
X_{12} = 1.07$\,MeV\,amu$^{-1}$ is the amount of energy needed to entrain one
atomic mass unit of $^{12}$C into the convection zone. Nuclear network \netone{}
releases $9.35$\,MeV from every single \reactCC{} reaction ($24$\,amu of
$^{12}$C), which is $q_\mrm{CC} = 0.390$\,Mev\,amu$^{-1}$. Since $q_\mrm{CC} <
\Gamma_\mrm{C12}$, the burning of the entrained material normally produces less
energy than what was needed to entrain it.\footnote{This also holds for nuclear
network \nettwo{}.} We have, however, increased the energy release from
\netone{} by the factor $f_\mrm{CC} = 10$ in run I11, so \mbox{$f_\mrm{CC}\,
q_\mrm{CC} = 3.90$\,MeV\,amu$^{-1} > \Gamma_\mrm{C12}$} holds for the amount of
energy produced per unit mass of fuel burnt in this run. More energy is produced
from the burning of the entrained material than what was needed to entrain it.
This closes a positive feedback loop that can explain the exponential growth
seen in Figs.~\ref{fig:luminosity_I11} and \ref{fig:entrainment_rate_I11}.

The main caveat of this hypothesis is that the entrainment rate is rather slow
to respond to changes in the total luminosity in other high-luminosity runs, see
Sect.~\ref{sect:entrainment_rate}. The $\varepsilon$ mechanism  \citep[see
e.g.][]{kippenhahn12} operating in the C-burning layer atop the convection zone
may also contribute to the development of large-amplitude oscillations at the
upper boundary, but the stability analysis involved in quantifying this
contribution is beyond the scope of this paper.

\section{Summary and conclusions}
\label{sect:summary}

We have employed 3D hydrodynamic simulations to investigate the dynamic feedback
from the burning of a C-rich material ingested from a stably-stratified layer
into a convective O-burning shell in a model of an evolved massive star. All but
one of the simulations reach a quasi-stationary state, in which the rates of
mass entrainment and burning are in a close balance. Most of our runs use a
$768^3$ Cartesian grid, but two runs using a $1152^3$ grid are in excellent
quantitative agreement with their $768^3$ counterparts
(Figs.~\ref{fig:entrainment_rate_vs_luminosity},
\ref{fig:L_C_fraction_vs_luminosity}, and \ref{fig:rms_velocity_vs_luminosity}).
This result provides further support for our previously established convergence
upon grid refinement for the same setup (J17) as well as for other simulations
performed using the \ppmstar{} code \citep{woodward15}.

The entrainment rate $\dot{M}_\mrm{e}$ is proportional to the total luminosity,
which in our suite of simulations spans almost three orders of magnitude (from
$3 \times 10^{10}$\,L$_\odot$ to ${\sim}10^{13}$\,L$_\odot$), although
$\dot{M}_\mrm{e}$ starts to deviate from that scaling in the lowest third of the
luminosity range. We suggest that these deviations might be caused by the
low-luminosity simulations' being too short for convection to erode the initial
transition layer between the two fluids and to form a new boundary consistent
with the properties of the convection and of the entrainment process.

Carbon burning contributes between $16\%$ and $33\%$ of the total luminosity in
the quasi-stationary state when only \reactCC{} reactions are considered.
\reactCO{} reactions turn out to be equally important when the luminosity is
high (${\sim}10^{13}$\,L$_\odot$) and they strongly dominate when the luminosity
is low ($\lesssim 10^{11}$\,L$_\odot$). The contribution of C burning to the
total luminosity is only ${\sim}15\%$ almost independently of the luminosity
when the network includes both \reactCC{} and \reactCO{} reactions. Because we
assumed the concentration of $^{12}$C in the fluid above the convection zone to
be five times higher than that in the C shell of the particular MESA model that
we started with, the feedback from C burning would likely be smaller still
without that increase. This is the case in the simulations of \citet{mocak18},
in which the concentration of $^{12}$C above their O shell ranges from
${\sim}10^{-5}$ to ${\sim}10^{-4}$.

The velocity field is dominated by spherical harmonic modes $l = 2, 3, 4$, i.e.\
there are only a few large convective cells that start the turbulent cascade.
The dominance of such global modes has been previously observed in simulations
providing a full-solid-angle view of convection in similar environments, e.g.
\citet{bazan_arnett94, porter_woodward00b, gilet13, jones17, muller17}. Mass
entrainment occurs where these large-scale flows run into one another while
turning over at the upper convective boundary and they start pulling slivers of
the C-rich fluid downwards. Full-sphere simulations are therefore necessary to
correctly quantify this process. The rms convective velocity scales with
$L_\mrm{tot}^{1/3}$ as expected and the magnitude of velocity at any given total
luminosity $L_\mrm{tot}$ agrees with simulations of J17 who did not include O
and C burning but instead used time-independent volume heating to drive
convection in the shell.

Large-scale asymmetries in the distribution of the entrained C-rich material
cause significant deviations of the burning rate from spherical symmetry. The
quadratic dependence of the \reactCC{} reaction rate on the mass fraction of
$^{12}$C enhances the C luminosity by up to a factor of two compared with an
estimate based on spherical averages when this reaction dominates C burning. On
the other hand, fluctuations in the density, pressure, and temperature are
smaller than $1\%$ in the convection zone proper. The only exception is run I11,
in which we experimentally increased the energy release from our simple
C-burning network by a factor of ten to explore the regime closer to a full,
dynamic-time-scale merger of the O and C shells. An exponentially-growing
instability starts immediately after the onset of C entrainment in I11 and
density fluctuations reach ${\sim}5\%$ in the convection zone when a dipolar
oscillation has developed in the flow field with a Mach number locally exceeding
$0.2$. Fluctuations of such magnitude are known to facilitate shock revival in
supernova explosion simulations \citep{couch_ott13, couch_ott15, muller_janka15,
muller17}. If such an instability occurs well before core collapse it might
power a SN impostor event assuming that there is a physical mechanism capable of
transporting a significant fraction of the kinetic energy contained in the O
shell (${\sim}3 \times 10^{47}$\,erg for $v_\mrm{rms} \sim 200$\,km\,s$^{-1}$)
to the star's extended envelope. The entrainment rate reached in the
instability, $2 \times 10^{-4}$\,M$_\odot$\,s$^{-1}$, is close to what is needed
to obtain significant production of the odd-Z elements Cl, K, and Sc according
to the nucleosynthesis calculations of \citet{ritter18}. Still, the energy
feedback from C ingestion in case I11 is smaller than the energy feedback
expected due to the maximum entrainment rate that may be encountered in a full,
dynamic O-C shell merger.

We show in Sect.~\ref{sect:stability} that the instability in run I11 could have
been caused by the fact that more energy was released from the burning of one
unit of entrained mass in I11 than what was needed to entrain that mass.
Although this is a direct consequence of our having increased the energy release
from C burning in this run, there are a number of energy producing reactions
that our simple nuclear network does not include, e.g.\ those caused by the
products of C burning or by the burning of $^{20}$Ne that is abundant in the C
shell. Our energy argument suggests that the instability would also occur with
less feedback from the burning if the entrainment rate was larger at the same
luminosity, which would likely happen if we considered an O shell with a softer
upper boundary. Finally, interaction between the convective flows in merging O
and C shells (the latter not considered in this work) could contribute to the
development of instabilities at the interface as well.

Animated visualisations of some of the simulations presented in this work are
available on Zenodo.org\footnote{\url{https://zenodo.org/record/2592134}} and on
YouTube\footnote{\url{https://goo.gl/8B7w29}}. Readers interested in analysing
the 3D data sets presented in this paper are encouraged to contact the authors
who can provide JupyterHub data exploration access via the
Astrohub\footnote{\url{https://astrohub.uvic.ca}} platform.

\section*{Acknowledgements}

RA, who completed most of this work as a CITA national fellow, acknowledges
support from the Canadian Institute for Theoretical Astrophysics and from the
Klaus Tschira Stiftung. PRW acknowledges NSF grants 1413548 and 1515792. FH
acknowledges support from a NSERC Discovery grant. This research was conducted
as part of the JINA Center for the Evolution of the Elements (NSF grant
PHY-1430152). NCSA's Blue Waters and Compute Canada/WestGrid provided the
computing and data processing resources for this project. We thank two anonymous
referees for their comments, which improved the quality of this paper. This work
benefited from the use of a large amount of free software, most importantly the
MESA stellar-evolution code, IPython/Jupyter notebooks, Python libraries
matplotlib and pyshtools, the FFmpeg software suite, Subversion revision control
system, or the LaTeX document preparation system.




\bibliographystyle{mnras}
\bibliography{c-ingestion}



\appendix
\renewcommand\thefigure{\thesection.\arabic{figure}}

\section{Luminosity curves}
\setcounter{figure}{0}

\begin{figure*}
\begin{minipage}{.4\textwidth}
  \centering
  \includegraphics[width=\textwidth]{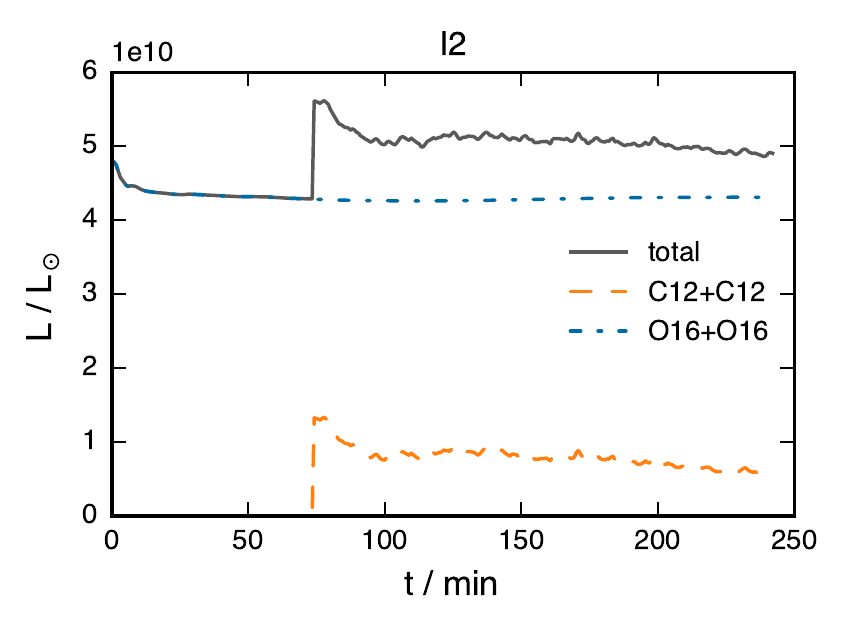}
\end{minipage}%
\begin{minipage}{.4\textwidth}
  \centering
  \includegraphics[width=\textwidth]{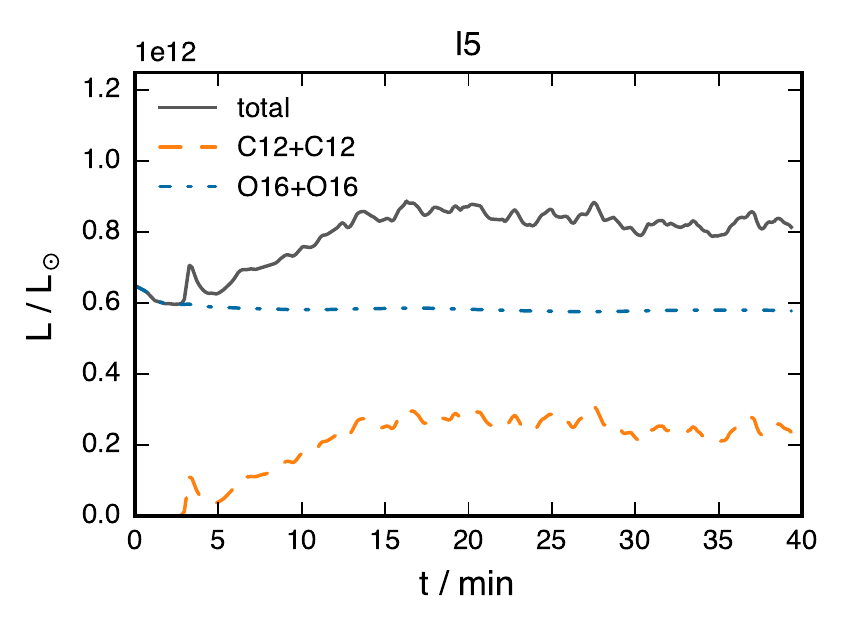}
\end{minipage}\\
\begin{minipage}{.4\textwidth}
  \centering
  \includegraphics[width=\textwidth]{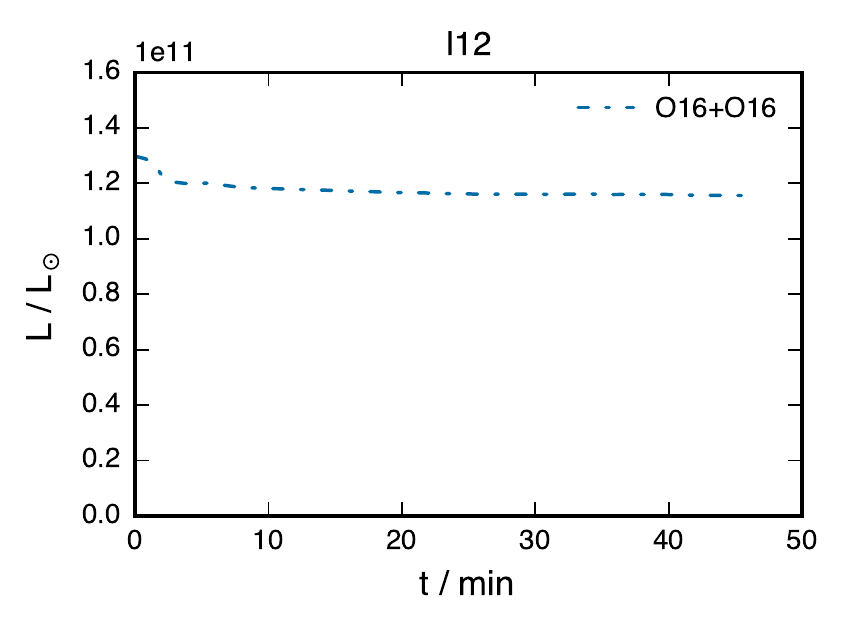}
\end{minipage}%
\begin{minipage}{.4\textwidth}
  \centering
  \includegraphics[width=\textwidth]{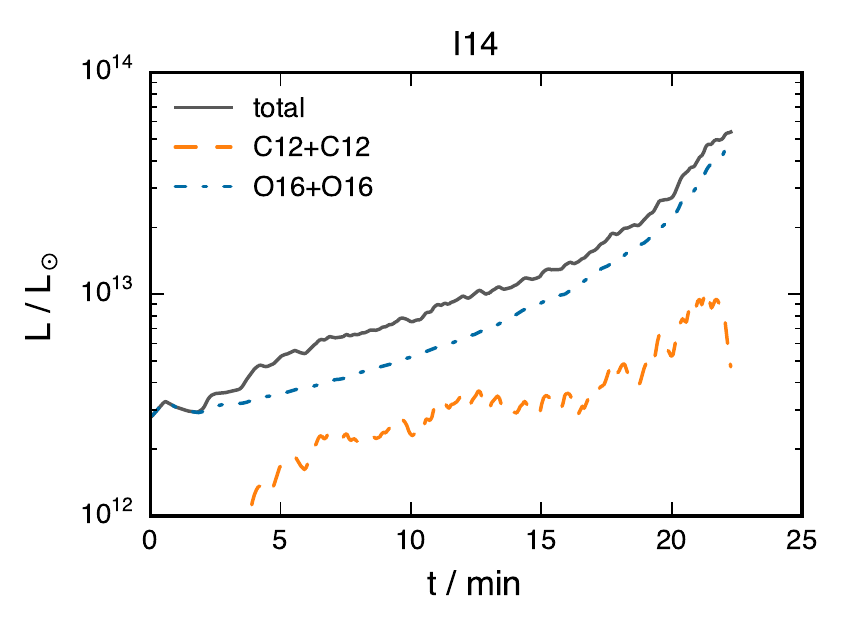}
\end{minipage}\\
\begin{minipage}{.4\textwidth}
  \centering
  \includegraphics[width=\textwidth]{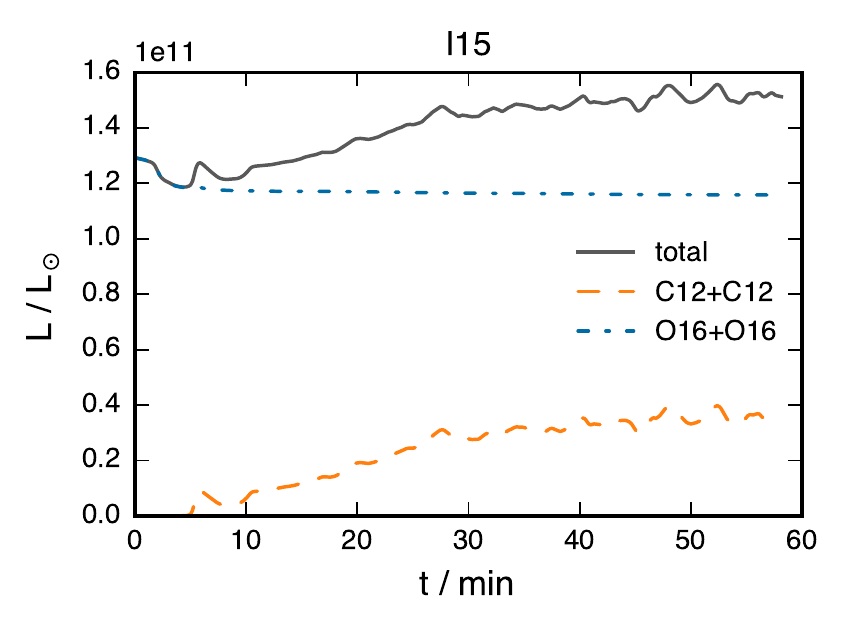}
\end{minipage}\\
\caption{As Fig.~\ref{fig:luminosity_evolution}, but runs I2 ($f_\mrm{OO} = 1$,
\netone{}), I5 ($f_\mrm{OO} = 13.5$, \netone{}), I12 ($f_\mrm{OO} = 2.7$, no C
burning), I14 ($f_\mrm{OO} = 67.5$, \netone{}), and I15 ($f_\mrm{OO} = 2.7$,
\netone{}) are shown.}
\label{fig:luminosity_evolution_appx}
\end{figure*}

\section{Entrainment rate measurements}
\setcounter{figure}{0}

\begin{figure*}
\begin{minipage}{.5\textwidth}
  \centering
  \includegraphics[width=\textwidth]{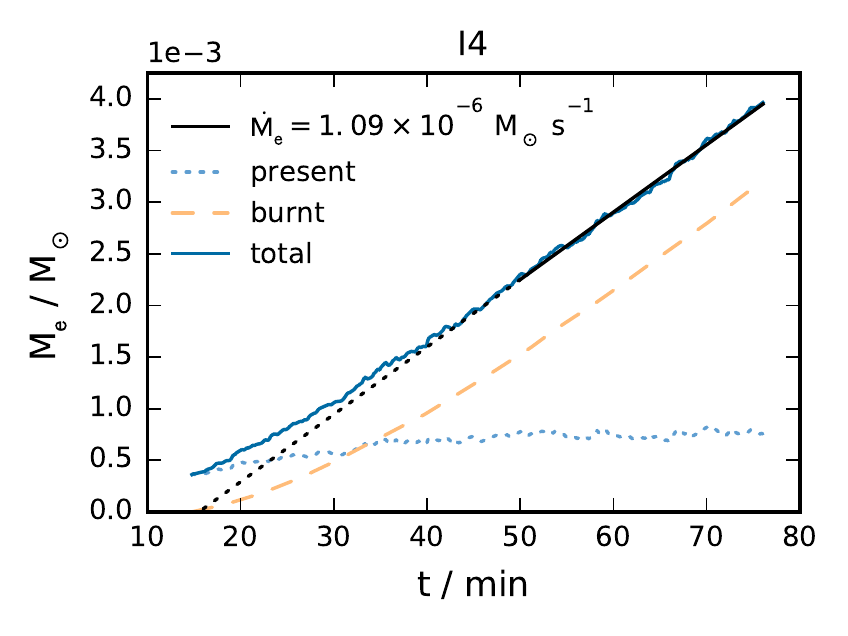}
\end{minipage}%
\begin{minipage}{.5\textwidth}
  \centering
  \includegraphics[width=\textwidth]{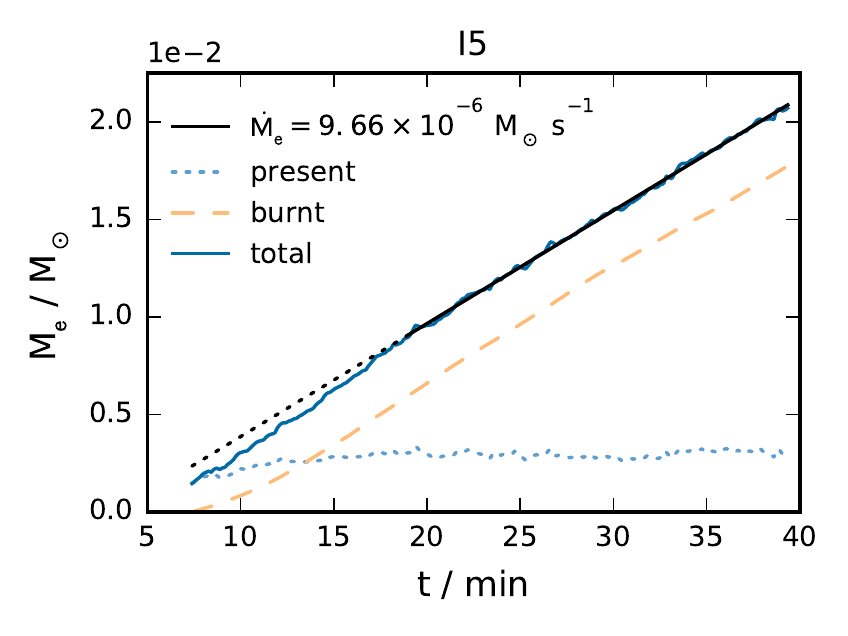}
\end{minipage}\\
\begin{minipage}{.5\textwidth}
  \centering
  \includegraphics[width=\textwidth]{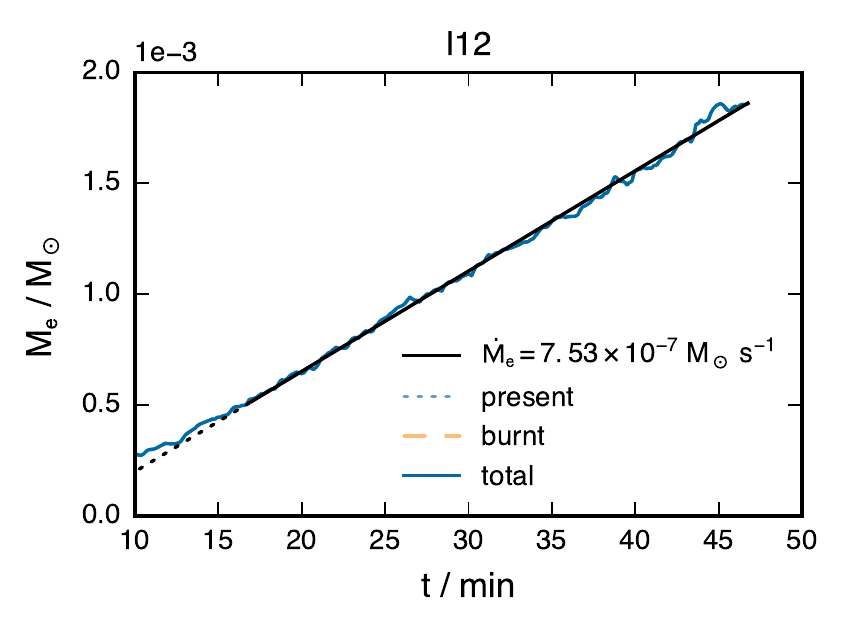}
\end{minipage}%
\begin{minipage}{.5\textwidth}
  \centering
  \includegraphics[width=\textwidth]{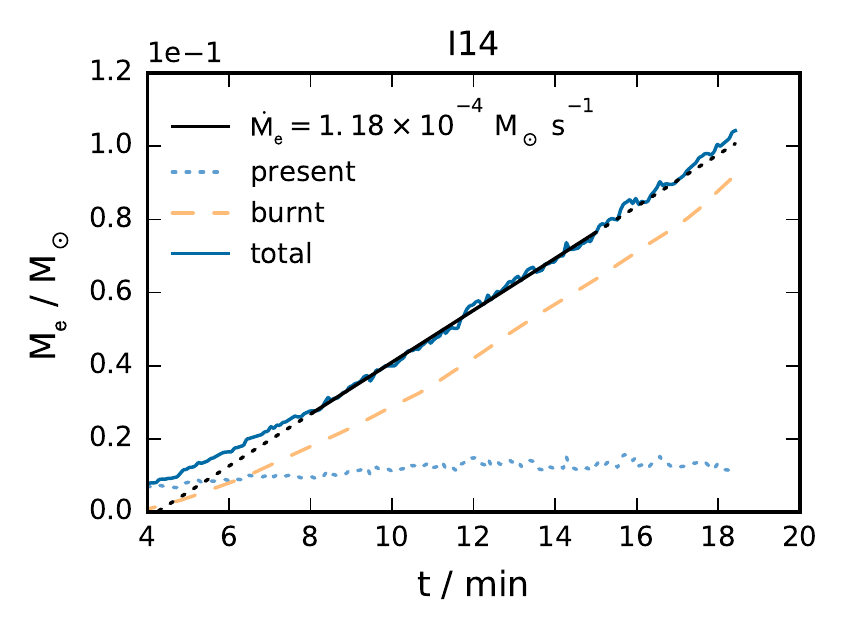}
\end{minipage}
\caption{As Fig.~\ref{fig:entrainment_rate}, but runs I4 ($f_\mrm{OO} = 2.7$,
\netone{}), I5 ($f_\mrm{OO} = 13.5$, \netone{}), I12 ($f_\mrm{OO} = 2.7$, no C
burning), and I14 ($f_\mrm{OO} = 67.5$, \netone{}) are shown.}
\label{fig:entrainment_rate_appx}
\end{figure*}

\begin{figure*}
\begin{minipage}{.5\textwidth}
  \centering
  \includegraphics[width=\textwidth]{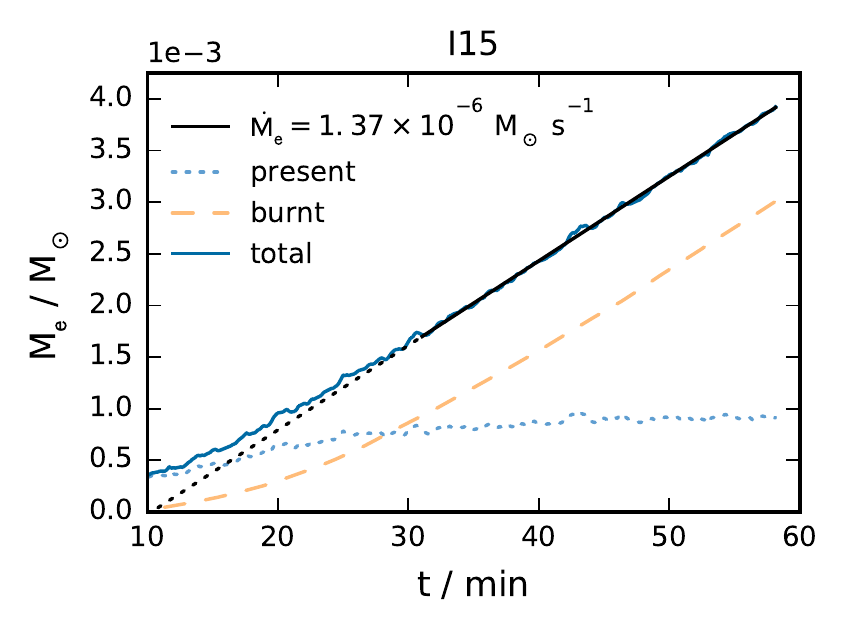}
\end{minipage}%
\begin{minipage}{.5\textwidth}
  \centering
  \includegraphics[width=\textwidth]{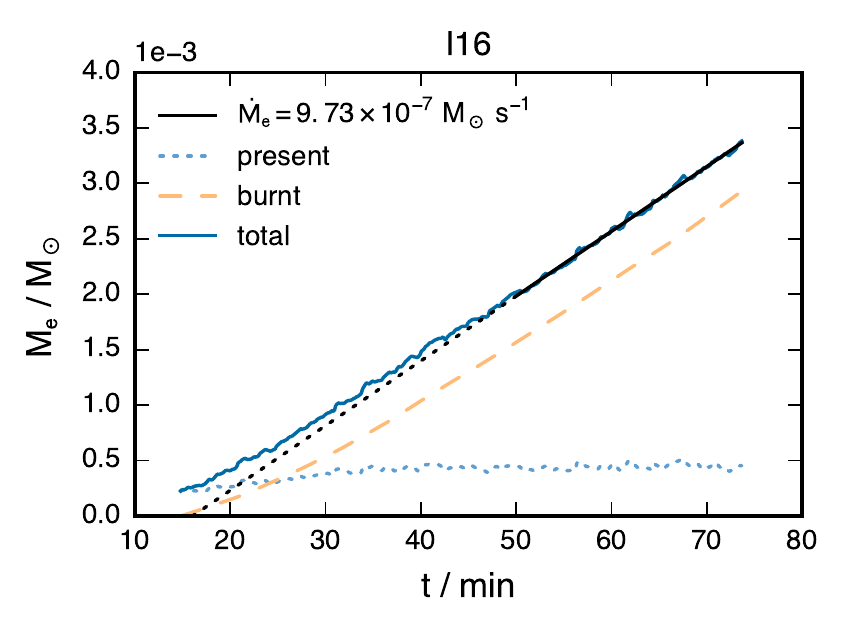}
\end{minipage}\\
\begin{minipage}{.5\textwidth}
  \centering
  \includegraphics[width=\textwidth]{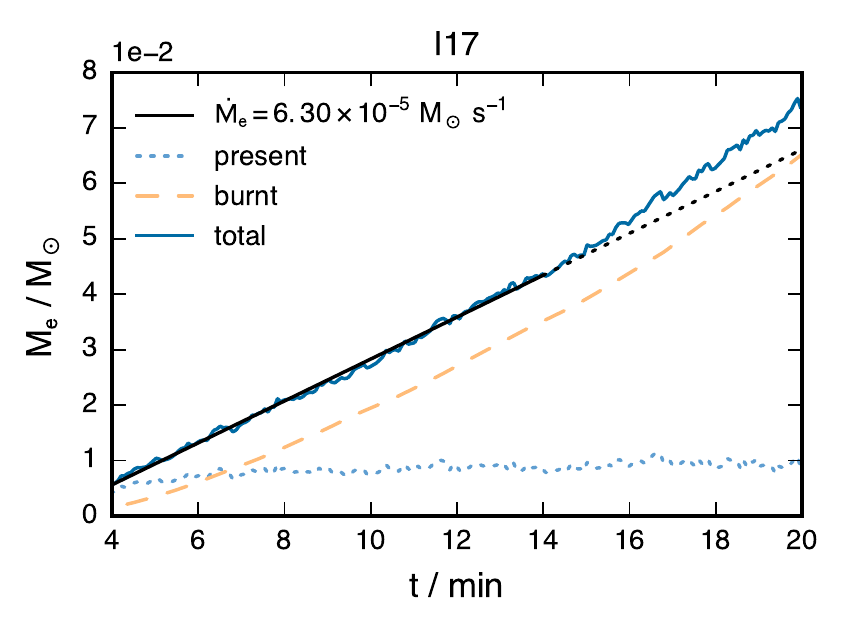}
\end{minipage}
\caption{As Fig.~\ref{fig:entrainment_rate}, but runs I15 ($f_\mrm{OO} = 2.7$,
\netone{}), I16 ($f_\mrm{OO} = 2.7$, \nettwo{}), and I17 ($f_\mrm{OO} = 67.5$,
\nettwo{}) are shown.}
\label{fig:entrainment_rate_appx2}
\end{figure*}

\section{Velocity power spectra}
\setcounter{figure}{0}

\begin{figure*}
\begin{minipage}{.5\textwidth}
  \centering
  \includegraphics[width=\textwidth]{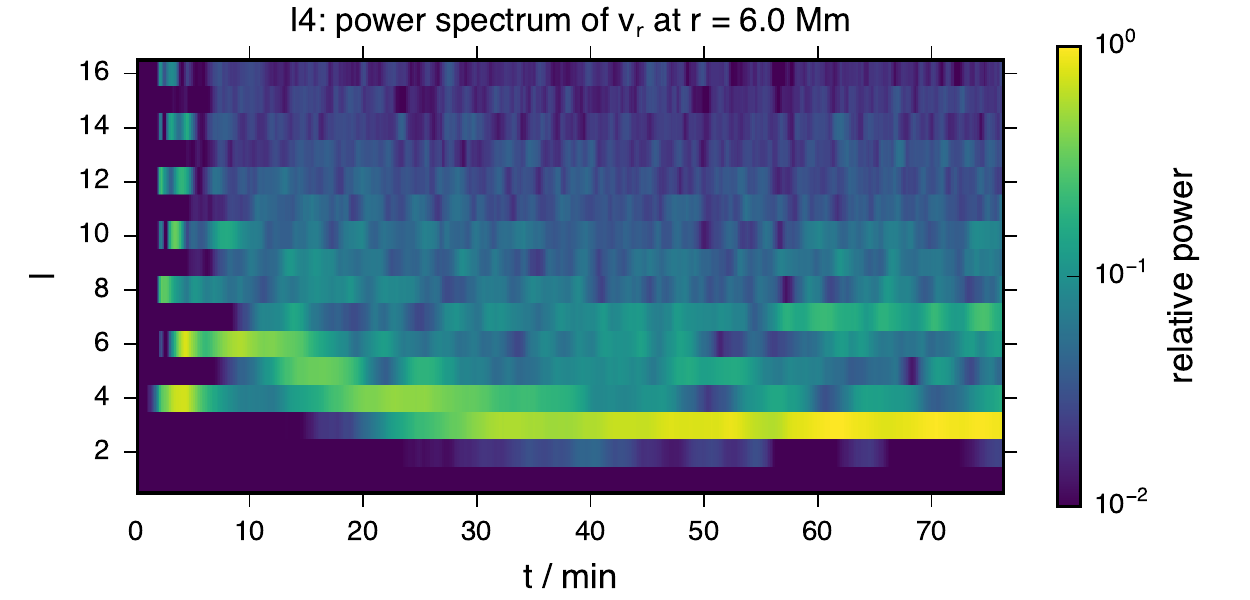}
\end{minipage}%
\begin{minipage}{.5\textwidth}
  \centering
  \includegraphics[width=\textwidth]{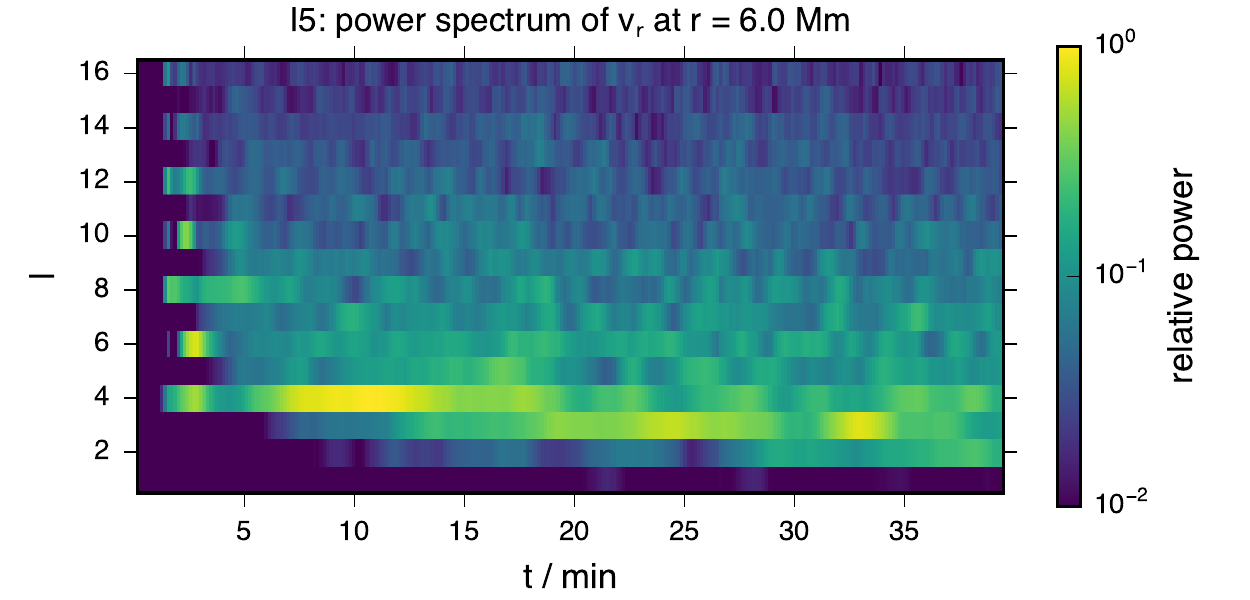}
\end{minipage}\\[1cm]
\begin{minipage}{.5\textwidth}
  \centering
  \includegraphics[width=\textwidth]{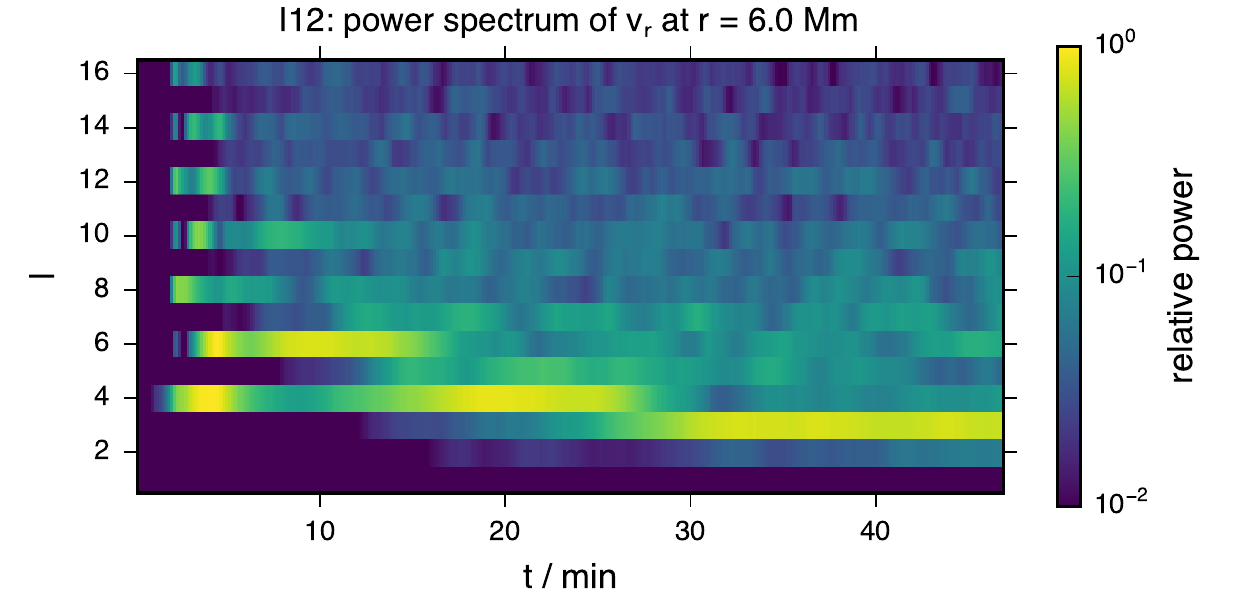}
\end{minipage}%
\begin{minipage}{.5\textwidth}
  \centering
  \includegraphics[width=\textwidth]{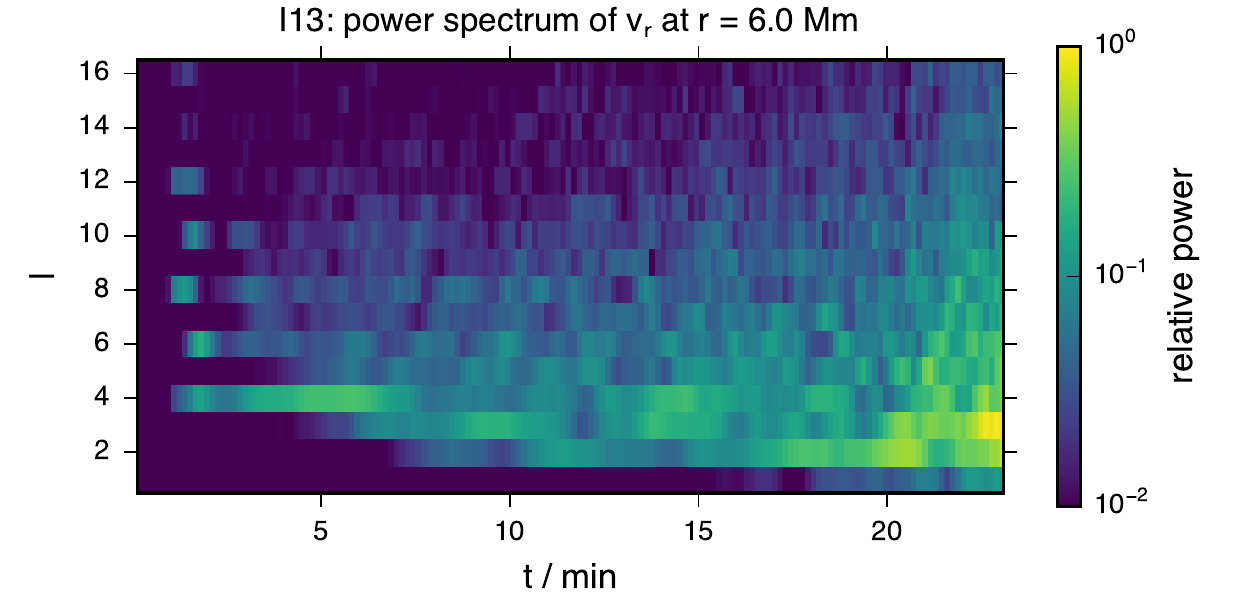}
\end{minipage}\\[1cm]
\begin{minipage}{.5\textwidth}
  \centering
  \includegraphics[width=\textwidth]{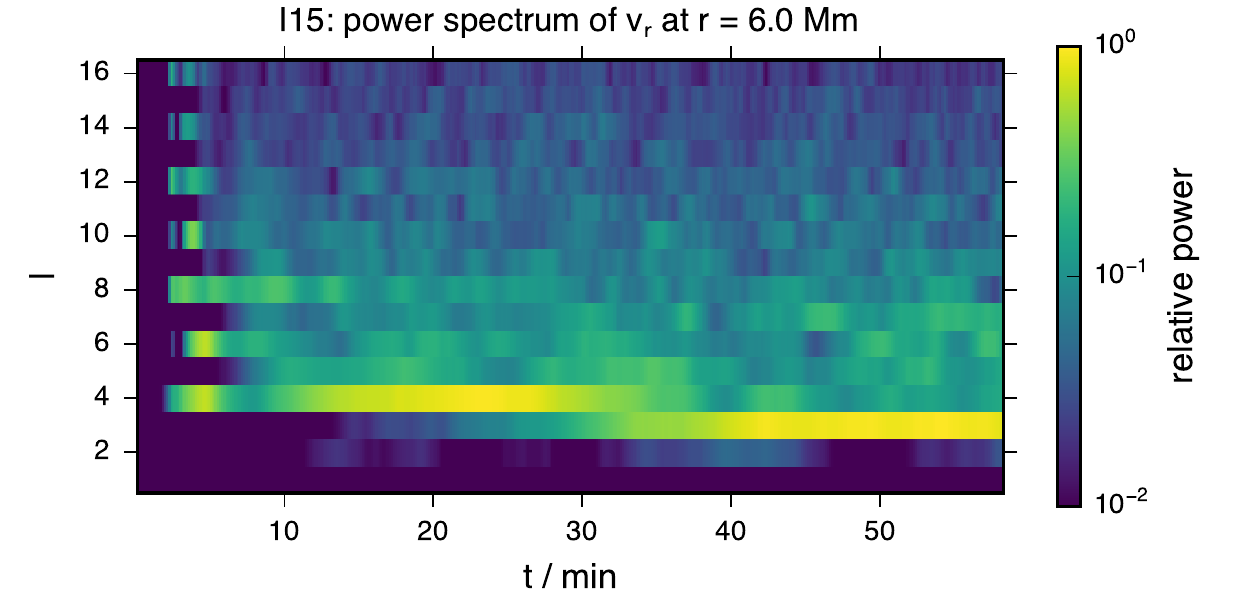}
\end{minipage}%
\begin{minipage}{.5\textwidth}
  \centering
  \includegraphics[width=\textwidth]{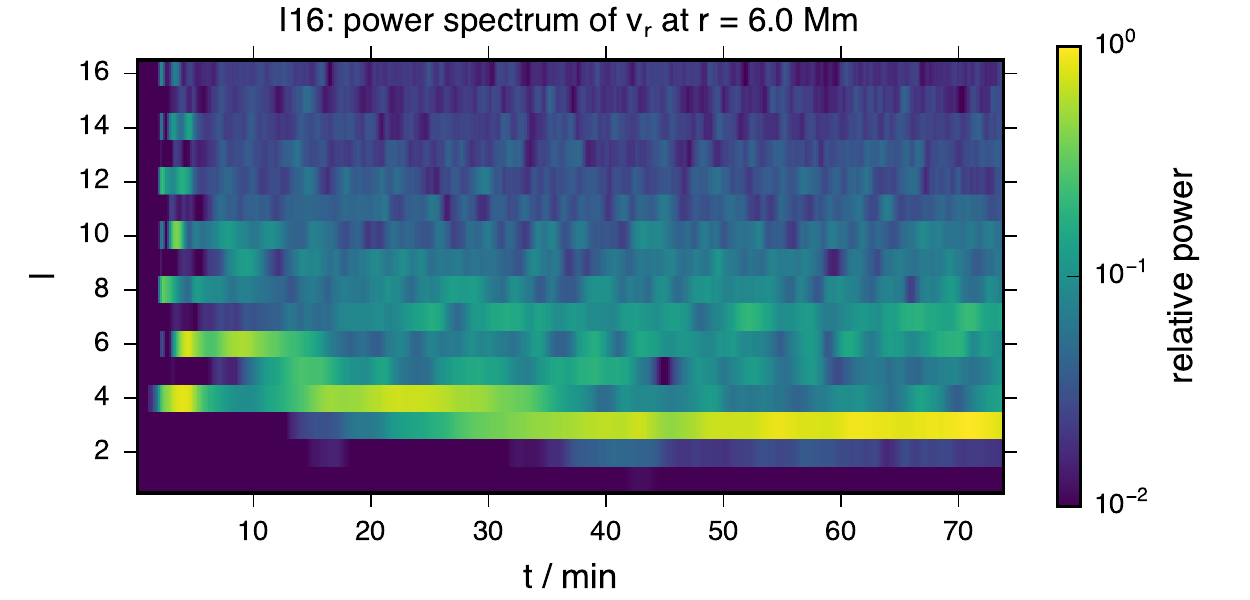}
\end{minipage}\\[1cm]
\begin{minipage}{.5\textwidth}
  \centering
  \includegraphics[width=\textwidth]{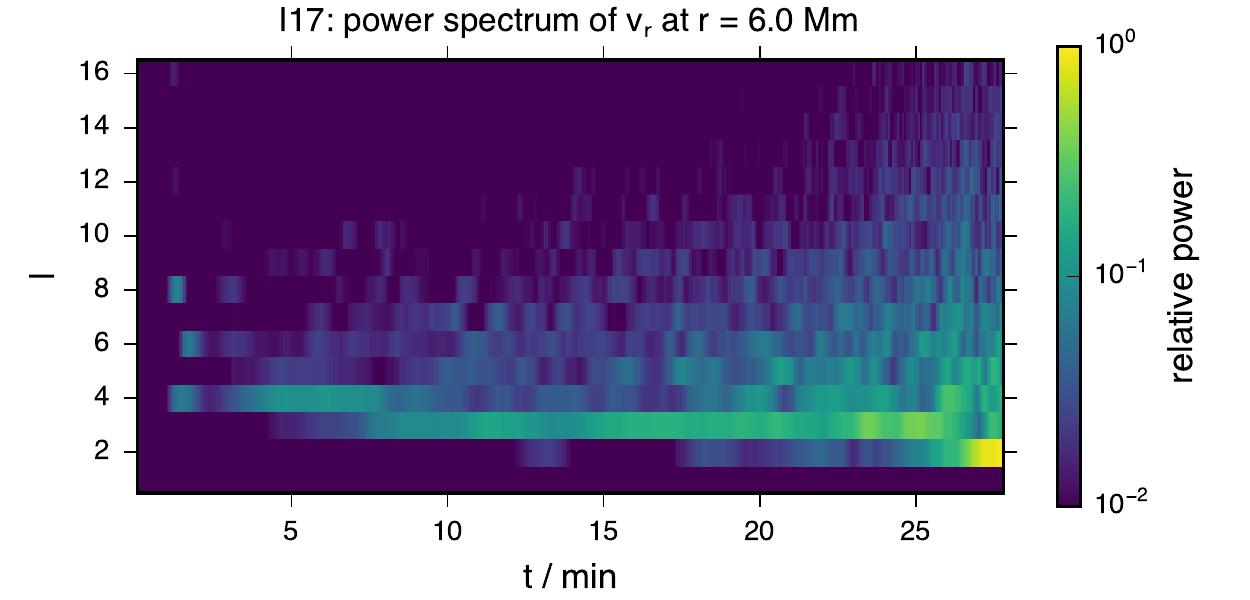}
\end{minipage}
\caption{As Fig.~\ref{fig:vr_spectrum_6.0_I14}, but runs I4 ($f_\mrm{OO} = 2.7$,
\netone{}), I5 ($f_\mrm{OO} = 13.5$, \netone{}), I12 ($f_\mrm{OO} = 2.7$, no C
burning), I13 ($f_\mrm{OO} = 67.5$, \netone{}), I15 ($f_\mrm{OO} = 2.7$,
\netone{}), I16 ($f_\mrm{OO} = 2.7$, \nettwo{}), and I17 ($f_\mrm{OO} = 67.5$,
\nettwo{}) are shown.}
\label{fig:vr_spectra_6.0_appx}
\end{figure*}


\bsp	
\label{lastpage}
\end{document}

%% file: macros/code.tex
\newcommand{\code}[1]{\texttt{#1}}

\newcommand{\mesa}{\code{MESA}}



%% file: macros/formatting.tex
\newcommand{\pan}[1]{{\textit{#1}}}

\newcommand{\FIGFF}[2]{{\ref{fig:#2}\pan{#1}}}

\newcommand{\FIG}[2]{{Fig.~\FIGFF{#1}{#2}}}
\newcommand{\Fig}[1]{{\FIG{}{#1}}}